\documentclass[12pt,a4paper,fleqn]{article} \textheight=25.5 true cm \textwidth=18 true cm
\usepackage{epsfig}
\usepackage{graphicx}
\usepackage{amsmath}
\usepackage{amssymb}
\usepackage{cite}
\usepackage{cleveref}
\usepackage{rotating}
\usepackage{pdfpages}
\begin{document}

\begin{center}
{\Large\bf \textit{Ab-initio} calculations on two-electron ions in strongly coupled plasma environment}\\[8 mm]
S. Bhattacharyya$^{1,*}$, J. K. Saha$^{2}$, T. K. Mukherjee$^{3}$ \\[4mm] 
$^{1}$\emph{Acharya Prafulla Chandra College, New Barrackpore, Kolkata 700131, India}\\
$^{2}$\emph{Indian Association for the Cultivation of Science, Jadavpur, Kolkata 700032, India}\\
$^{3}$\emph{Narula Institute of Technology, Agarpara, Kolkata 700109, India} \\
\vspace{0.5 cm}
*E-mail : sukhamoy.b@gmail.com
\end{center}
\vspace{0.5 cm}
\begin{abstract}
In this work, the controversy between the interpretations of recent measurements on dense aluminum plasma created with Linac coherent light sources (LCLS) X-ray free electron laser (FEL) and Orion laser has been addressed. In both kind of experiments, helium-like and hydrogen-like spectral lines are used for plasma diagnostics . However, there exist no precise theoretical calculations for He-like ions within dense plasma environment. The strong need for an accurate theoretical estimates for spectral properties of He-like ions in strongly coupled plasma environment leads us to perform \textit{ab initio} calculations in the framework of Rayleigh-Ritz variation principle in Hylleraas coordinates where ion-sphere potential is used. An approach to resolve the long-drawn problem of numerical instability for  evaluating two-electron integrals with extended basis inside a finite domain is presented here. The present values of electron densities corresponding to disappearance of different spectral lines obtained within the framework of ion-sphere potential show excellent agreement with Orion laser experiments in \textit{Al} plasma and with recent theories. Moreover, this method is extended to predict the critical plasma densities at which the spectral lines of H-like and He-like carbon and argon ions disappear. \textit{Incidental degeneracy} and \textit{level-crossing} phenomena are being reported for the first time for two-electron ions embedded in strongly coupled plasma. Thermodynamic pressure experienced by the ions in their respective ground states inside the ion-spheres are also reported.  
\end{abstract}

\vspace{0.8cm}
\section{Introduction}
The study of confined quantum mechanical systems has attracted immense attention from researchers around the world due to the novel and unusual structural properties exhibited by such systems when subject to spatial limitation \cite{aqc1}. A wide variety of physical situations are manifested in nature that relates to spatially confined systems such as atoms or molecules trapped in zeolite sieves \cite{jac2}, fullerenes \cite{xu3}, plasma environment \cite{sil5}, solvent environment \cite{can6}, under high pressure in the walls of nuclear reactors \cite{wal7}, quantum dot or artificial atom \cite{kos8}, molecular containers, storage of fuel cells \cite{cio9,tur10}, matter under high pressure in Zovian planets \cite{gui11} \textit{etc}. Along with the experimental and technological development, theoretical research plays a fundamental role for designating appropriate models in order to explore and predict the behavioral changes of a confined system. The present study is focused on atomic systems embedded in plasma environment. In recent years, atoms placed in external plasma environment have received considerable attention from researchers \cite{sah12,jqs13,jpb14,ord16,ho17,das18,sbz19,das20,san15} due to their wide applications in various disciplines of science \textit{e.g.} astrophysics, condensed matter physics, biology \textit{etc}. While dealing with plasma that follows classical statistics, a coupling parameter ($\Gamma$) defined as the ratio of the average electrostatic energy and the average thermal energy is introduced. $\Gamma < 1$ corresponds to weakly coupled plasma (WCP) for which the effective potential experienced by the embedded ion is expressed according to \textit{Debye} model \cite{akh21} and $\Gamma \geq 1$ denotes strongly coupled plasma (SCP) where the potential is taken from Ion-sphere (IS) model \cite{ich22}. According to the IS model, a sphere (termed as \textit{Wigner-Seitz sphere}) surrounding a positively charged ion is considered in such a way that the plasma electrons within the sphere neutralize the positive ion. The size of the \textit{Wigner-Seitz} sphere will decrease when the number density of plasma electrons ($n_e$) increases. The temperature ($T$) of the plasma does not appear directly in this model but it is implicit as $n_e$ is different for different temperatures. The domain of the effective potential representing the SCP surrounding is finite in case of IS model in contrast to the long range character of the screened Coulomb potential used in \textit{Debye} model \cite{akh21} for WCP environment. The examples of WCP's are the gaseous discharge plasma ($T \sim 10^{4}$ $K$ and $n \sim 10^{11} $/c.c), plasma in controlled thermo-nuclear reaction ($T \sim 10^{8}$ $K$ and $n \sim 10^{16}$/c.c), solar coronal plasma ($T \sim 10^{6}-10^{8}$ $K$ and $n \sim 10^{6}-10^{10}$/c.c), Tokamak plasma ($T \sim 10^{5}-10^{7}$ $K$ and $n \sim 10^{8}-10^{16}$/c.c) etc. SCP's (temperature varies and typical densities $\geq$ $10^{23}$/c.c.) are observed in highly evolved stars in high density states, interior of Jovian planets, explosive shock tubes, two-dimensional states of electrons trapped in surface states of liquid helium, laser produced plasmas etc. Spectral line shifts, pressure ionization, ionization potential depression (IPD) and line merging phenomena occur in SCP environment due to the deformation of the ionic potential by the plasma fields. Such properties and knowledge about ion-plasma interaction can effectively be utilized for diagnostics and the investigation of X-ray opacity of matter under conditions prevailing in stellar interiors. The experimental observations using laser produced plasmas for $C$, $Al$ and $Ar$ by Nantel \textit{et. al.} \cite{nan23}, Saemann \textit{et. al.} \cite{sae24} and Woolsey \textit{et. al.} \cite{woo26} have explicitly demonstrated the effect of SCP on the spectral properties of such systems. The laboratory plasma conditions ($T$ and $n_e$) undergo rapid changes \textit{w.r.t.} where local thermodynamic equilibrium is not maintained. Consequently, the experimental measurements become extremely complicated leading to a loss of accuracy and, till the end of the last century, this accuracy level was not even mentioned in most of the experiments. \\\\
In recent years, a remarkable improvement has been made \cite{vin,vin1,ciri,cho} with the advent of Linac Coherent Light Sources (LCLS) towards creation of relatively long-lived high-density plasma at homogeneous temperature and densities. In these experiments, X-ray free-electron Laser (FEL) was used to create plasma with densities up to almost one order higher than solid \textit{Al} and then spectral line profiles of different charge states of \textit{Al} were used for diagnostics. The effect of IPD on the emitted spectra as a function of $n_e$ is explored experimentally by observing the disappearance of spectral lines of H-like and He-like \textit{Al}. During the observation of K-shell fluorescence of highly charged \textit{Al}, Ciricosta \textit{et al.} \cite{ciri} found that the IPDs measured were not consistent with the predictions of the most widely used theoretical model of Stewart and Pyatt (SP) \cite{ste28} but in good agreement with an earlier model due to Ecker and Kr{\"o}ll (EK) \cite{ek}. However, this observation was questioned in a subsequent theoretical study by Preston \textit{et al.} \cite{pre} where detailed simulations were carried out for the spectral lines of H-like and He-like \textit{Al} to study IPD by using both SP and EK (in a modified form) models. In experiments, the intensities and Stark-broadened widths of He-$\beta$ and Ly-$\beta$ spectral lines are used for main diagnostics. A direct measurement of ionization potential depression is a difficult task because of its indistinguishability from the effect of spectral line merging due to Stark broadening \cite{ing}.  Hoarty \textit{et al.} \cite{hoa,hoa1} have been able to overcome this difficulty and their measurements for \textit{Al} plasma using Orion laser are in closer agreement with SP model of IPD than the EK model. This situation clearly warrants extensive and accurate \textit{ab initio} study of atomic structures within dense plasma environment. Very recently, Son \textit{et al.} \cite{son} have adopted a two step Hartree-Fock-Slater approach to assess the IPD effect for $ Al^{3+} $ to $Al^{7+}$ within plasma where a muffin-tin flat potential was used. The IPDs calculated by Son \textit{et al.} \cite{son} lie between the SP and modified EK models and in some cases, are close to SP model. But so far, no extensive theoretical calculation on IPDs for He-like ions has been performed. It should be noted here that both SP model and EK model for estimating IPDs are derived within the framework of IS potential. The only theoretical work for He-like ions in the field of SCP by using IS potential is due to Sil \textit{et. al.} \cite{pkm27} where both non-relativistic and relativistic calculations were carried out using time-dependent perturbation theory. Though Sil \textit{et. al.} \cite{pkm27} demonstrated that the relativistic IS model yields consistent results in predicting the spectral line positions for the systems considered, some anomalies such as better agreement of non-relativistic results with experiments than relativistic ones are observed in their data \cite{pkm27}. Such strange features may arise due to improper inclusion of electron correlation in basis set within a finite region. A major challenge for precise theoretical calculations is, therefore, to develop an appropriate methodology where the effect of electron correlations within a finite domain is aptly included.\\\\
To the best of our knowledge, there exists no calculation of He-like atoms embedded in SCP using Hylleraas type basis set though it is well accepted that within the framework of Ritz variational technique, explicitly correlated wave functions expanded in terms of Hylleraas basis (and its variants) can produce most accurate non-relativistic energies of He-like atoms. These methods have been applied extensively to free He-like systems whereas for spatially confined two-electron systems, such studies are limited to S states only \cite{aqun1,flo2,lau35,mont1}. According to Laughlin and Chu \cite{lau35}, the generalized Hylleraas basis sets used in such calculations suffer the loss of linear independence for large dimensions of the wave functions and hence all the calculations \cite{aqun1,flo2,lau35,mont1} were limited to small dimensions (at best 25). Laughlin and Chu \cite{lau35} made an effort to remove this difficulty and extended the basis size up to 95 parameters where they have to compromise with the flexibility of the non-linear parameters. Recently, for $^1S^e$ states of He-like systems under spherical confinement, the present authors have calculated the energy values \cite{scr33} by using standard Hylleraas basis set of dimension 161 and the results have been confirmed by Montgomery and Pupyshev \cite{mont2}. In the present work, a successful effort has been made to develop a general methodology in Hylleraas basis for both \textit{S} and \textit{P} states of He-like systems. The finite domain two-electron integrals with flexible parameters are evaluated where the problem of linear dependency in larger dimensions is clearly avoided.\\\\
We have estimated precise non-relativistic energy values of $1sns$ ($^{1}S^{e}$) [$n=1-3$] and $1sn'p$ ($^{1}P^{o}$) [$n'=2-4$] states of He-like  $C$, $Al$ and $Ar$ within SCP environment. Accuracy of the computed energy eigenvalues have been tested systematically over an extended range of parameters and also by increasing the number of terms ($N$) in the expanded basis sets. The plasma densities ($n_{e}$) are varied from a low value that corresponds to almost a free system to a very high one that leads the ion towards destabilization. The plasma electron densities in different experimental conditions \cite{nan23,sae24,woo26,vin,vin1,ciri,cho,hoa} are well covered within the density ranges studied here. The energy eigenvalues of $ns$ ($^{2}S$) [$n=1-2$] and $n'p$ ($^{2}P$) [$n'=2-3$] states of H-like $C$, $Al$ and $Ar$ in SCP are also estimated to determine the variation of ionization potential (IP) \textit{w.r.t.} $n_{e}$. As $n_{e}$ increases, both the two-electron excited states as well as the respective one-electron threshold move towards destabilization, thereby reducing the IP. It is remarkable that after a certain value of $n_{e}$, the two-electron energy levels move above the respective one-electron energy level and become quasi-bound. \textit{Incidental degeneracy} \cite{sen38} and subsequent \textit{level-crossing} phenomenon between the excited states such as $1s2s$ ($^{1}S^{e}$) and $1s2p$ ($^{1}P^{o}$) under SCP have been observed. Such features are novel in the context of foreign atoms in SCP and being reported for the first time in literature. Due to spatial restriction imposed upon the wave function according to IS model under SCP environment, the ion will feel a pressure inside the \textit{Wigner-Seitz} sphere. The variation of thermodynamic pressure \textit{w.r.t.} plasma density is also calculated. The paper is organized as: an outline of the basic theory used and details on the evaluation of the basis integrals are given in Section 2, followed by a discussion on the results in Section 3 and finally concluded in section 4 with a view towards further application of the present methodology in related fields.

\section{Method}
The non-relativistic Hamiltonian (in a.u.) of a two-electron ion placed inside SCP environment can be written as
\begin{eqnarray}\label{1}
H=\sum^{2}_{i=1}\left[-\frac{1}{2}\nabla_{i}^{2}+V_{IS}(r_{i})\right]+\frac{1}{r_{12}}
\end{eqnarray}
$V_{IS} (r_i)$ is the one-electron term of the modified potential energy as `$seen$' by the \textit{i-th} electron within plasma environment. It is to be noted that in this model, the electronic repulsion part in the potential is completely unaltered. The spherically symmetric potential $V_{IS} (r_i)$ experienced by a positive charge ion surrounded by a one-component plasma within the ion-sphere \cite{ich22} is given by
\begin{eqnarray}\label{2}
V_{IS} (r_i)&=&-\frac{Z}{r_i}+\frac{(Z-N_e)}{2R}\left[3-{\left(\frac{r_i}{R}\right)}^2\right]
\end{eqnarray}
where \textit{R} is the \textit{Wigner-Seitz} radius \cite{ich22}, \textit{i.e.} the radius of the surrounding ion-sphere, $Z$ is the nuclear charge and $N_e$ ($<Z$) is the number of electrons present in the ion. For helium-like ions, $N_e=2$ is being taken. The Schrodinger equation $H\Psi=E\Psi$ is to be solved to obtain the energy eigenvalues where the wave function is subject to the normalization condition $\langle \Psi \vert \Psi \rangle=1$ within the sphere. The structure of the potential demands that there is no electron current taking place through the boundary surface of \textit{Wigner-Seitz} sphere, and the orbital wave function $\Psi$ satisfies the boundary condition 
\begin{eqnarray}\label{eq:ab}
\Psi(r)=0~~~~~~~~~~~~~at~~~ r \geq R
\end{eqnarray}
This boundary condition plays a significant role in behavioral changes of the confined atoms. The plasma electrons within the ion-sphere neutralize the central positive charge and the size of the \textit{Wigner-Seitz} sphere is determined by the condition of overall charge neutrality that yields 
\begin{eqnarray}\label{3}
R={\left[\frac{3(Z-N_e)}{4\pi n_e}\right]}^\frac{1}{3}
\end{eqnarray}
The above expression for `$R$' is used to determine the IPD according to the SP model \cite{ste28}. However, in EK model \cite{ek} for determining the IPD, this radius was calculated in a somewhat different way where both the electron density ($n_e$) and ion density ($n_i$) are considered. According to EK model \cite{ek}, the radius of the sphere would be expressed as
\begin{eqnarray}\label{3}
R_{EK}={\left[\frac{3}{4\pi (n_e+n_i)}\right]}^\frac{1}{3}
\end{eqnarray}
The EK model \cite{ek} is relevant when the ion density is appreciably high and can affect the mean separation of the free electrons.\\\\
Due to the translational symmetry of the Hamiltonian of a two-electron ion, the degrees of freedom reduce from nine to six by separating the motion of the centre of mass. These six coordinates can be taken as the sides of the triangle $r_1$, $r_2$, $r_{12}$ formed by the three particles, \textit{i.e.}, two electrons and the fixed nucleus and the Eulerian angles $(\theta,\phi,\psi)$ defining the orientation of this triangle in space. 
The wave function obeying symmetry properties under particle exchange may be written as \cite{bha31}
\begin{eqnarray}\label{eq:wf}
\Psi\left(\overrightarrow{r_1},\overrightarrow{r_2}\right)=\sum_{\kappa}\left[f_L^{\kappa +}\left(r_1,r_2,\theta_{12}\right)D_L^{\kappa +}(\theta,\phi,\psi)+f_L^{\kappa -}\left(r_1,r_2,\theta_{12}\right)D_L^{\kappa -}(\theta,\phi,\psi)\right]
\end{eqnarray}
$\theta_{12}$ is the angle between $\overrightarrow{r_1}$ and $\overrightarrow{r_2}$. The summation in eq. (6) goes over every alternate value of $\kappa$, where $\kappa = |k|$. $k$ is the angular momentum quantum number about the body fixed axis of rotation whose value satisfies $k \leqslant L$, $L$ being the total angular momentum quantum number. The symmetric top functions $D_L^{\kappa +}$ and $D_L^{\kappa -}$ are the eigenfunctions of angular momentum operator $L^2$ of the two electrons. The rotational invariance of the Hamiltonian makes it possible to express the variational equation of two electrons in the field of a fixed nucleus in terms of three independent variables $r_1, ~r_2$ and $r_{12}$ (or $\theta_{12}$). The reduction of the Eulerian angles from the variational equation is an immediate consequence of the spherical symmetry of the field.
\subsubsection*{For $^{1}S^{e}$ state} 
The variational equation for $^{1}S^{e}$ states originating from two s-electrons (1$s$n$s$ configuration) is given by, following ref. \cite{tkm32},
\begin{eqnarray*}\label{eq:ab1}
\delta\int \left[\frac {1}{2}\left(\frac{d\Psi_S}{dr_1}\right)^2+ \frac {1}{2}\left(\frac{d\Psi_S}{dr_2}\right)^2+\left(\frac{r_1^2-r_2^2+r_{12}^2}{2r_1r_{12}}\right)\left(\frac{d\Psi_S}{dr_1}\right)\left(\frac{d\Psi_S}{dr_{12}}\right)\right.\nonumber~~~
\end{eqnarray*}
\begin{equation}
\left.+\left(\frac{r_2^2-r_1^2+r_{12}^2}{2r_2r_{12}}\right)\left(\frac{d\Psi_S}{dr_2}\right)\left(\frac{d\Psi_S}{dr_{12}}\right)+(V-E)\Psi_S^2\right]dV_{r_1,r_2,r_{12}}~=~0
\end{equation}
where the upper limit of integration for $r_1$ and $r_2$ is $R$ in contrast to infinity for the free atomic case. The upper and lower limits of integration for $r_{12}$ are $(r_1 + r_2)$ and $|r_1 - r_2|$, respectively. 
The volume element is expressed as 
\begin{eqnarray}
dV_{r_1,r_2,r_{12}}~=~r_1 r_2 r_{12} dr_1 dr_2 dr_{12}
\end{eqnarray}
For $S$-states, $L=\kappa=0$ and the wave function $\Psi_S$ can be written as
\begin{eqnarray}\label{eq:n1}
\Psi_S=f_0^0 D_0^0 = f_S + \tilde{f}_S
\end{eqnarray} 
where $\tilde{f}_S (r_1, r_2, r_{12})=f_S (r_2, r_1, r_{12})$. The correlated wave function \cite{scr33} is written as 
\begin{eqnarray}\label{eq:ab2}
f_S (r_1,r_2,r_{12})=(R-r_1)(R-r_2)f(r_1,r_2,r_{12})
\end{eqnarray}
with
\begin{eqnarray}\label{eq:ab2}
f(r_1,r_2,r_{12})=e^{-\sigma_1 r_1-\sigma_2 r_2}\sum_{l\geq 0}\sum_{m\geq 0}\sum_{n\geq 0}C_{lmn}r_1^l r_2^m r_{12}^n
\end{eqnarray}
$\sigma_1$ and $\sigma_2$ are the nonlinear parameters taking care of the effect of radial correlation in the wave function whereas the angular correlation effect is incorporated through different powers of $r_{12}$. $C$'s are the linear variational parameters. The total number of parameters (\textit{N}) in the basis set is defined as the total number of different $(l, m, n)$ sets (eq. 11) taken in the expansion of $f(r_1,r_2,r_{12})$.
\subsubsection*{For $^{1}P^{o}$ state}
For $P$ state of odd parity $(L=1$ and $\kappa=+1, -1)$, the total wave function can be written as \cite{cpl34}
\begin{equation}
\Psi_P=f^{1+}_{1}D^{1+}_{1}+f^{1-}_{1}D^{1-}_{1}
\end{equation}
The origin of $^{1}P^{o}$ state due to $sp$ configuration of two-electron atoms can be shown by expressing $D^{1+}_{1}$ in terms of individual polar coordinates $(\theta_1,\phi_1;\theta_2,\phi_2)$ as given below:
\begin{equation}
D^{1+}_{1}(\theta,\phi,\psi)\cos\frac{\theta_{12}}{2}=P_0^0(\cos\theta_1)P_1^0(\cos\theta_2)+interchange
\end{equation}
Similar expressions for $D^{1-}_{1}$ can be derived. 
After integration over the Eulerian angles, the variational equation for $^{1}P^{o}$ states originating from 1$s$n$p$ configuration reduces to \cite{cpl34},
\begin{eqnarray*}
\delta \int\left[\sum_{i=1}^2 \left\{\left({{\partial f_1^{1 +}}\over{\partial r_i}}\right)^2 +\left({{\partial f_1^{1 -}}\over{\partial r_i}}\right)^2\right\}+\left(\frac{1}{r_1^2}+\frac{1}{r_2^2}\right)\left\{\left({{\partial f_1^{1+}}\over{\partial\theta_{12}}}\right)^2+\left({{\partial f_1^{1-}}\over{\partial \theta_{12}}}\right)^2+\left(\frac{1}{4}+\frac{1}{2\sin^2\theta_{12}}\right)\times \right.\right. \nonumber
\end{eqnarray*}
\begin{equation}
\left.\left.  \left\{\left(f_1^{1+}\right)^2+\left(f_1^{1-}\right)^2\right\} + \frac{\cos\theta_{12}}{2\sin^2\theta_{12}}\left\{\left(f_1^{1+}\right)^2-\left(f_1^{1-}\right)^2\right\}\right\}+ \left(\frac{1}{r_2^{2}}-\frac{1}{r_1^{2}}\right)\left\{\left({{\partial f_1^{1+}}\over{\partial \theta_{12}}}f_1^{1-}-{{\partial f_1^{1-}}\over{\partial \theta_{12}}}f_1^{1+}\right) \right.\right.\nonumber~~~~~~~~~~~~
\end{equation}
\begin{equation}
\left.\left.-\frac{1}{\sin\theta_{12}}f_1^{+1}f_1^{-1}\right\}+2(V-E)\left\{\left(f_1^{1 +}\right)^2+ \left(f_1^{1-}\right)^2\right \}\right]dV_{r_1,r_2,\theta_{12}}=0~~~~~~~~~~~~~~~~~~~~~~~~~~~~~~~~~~~~~~~~~~~~~~~~~~~~~~~~~~~~~~~~~~~
\end{equation}
The correlated functions $f_1^{1+}$ and $f_1^{1-}$ can be written as
\begin{equation}
f^{1+}_{1}=(f_P+\tilde{f}_P)\cos\frac{\theta_{12}}{2}
\end{equation}
\begin{equation}
f^{1-}_{1}=(f_P-\tilde{f}_P)\sin\frac{\theta_{12}}{2}
\end{equation}
where
\begin{eqnarray}\label{eq:ab222}
f_P(r_1,r_2,r_{12})=(R-r_1)(R-r_2)g(r_1,r_2,r_{12})
\end{eqnarray}
and
\begin{eqnarray}\label{eq:ab2}
g(r_1,r_2,r_{12})=e^{-\rho_1 r_1-\rho_2 r_2}\sum_{l\geq 1}\sum_{m\geq 0}\sum_{n\geq 0}D_{lmn}r_1^l r_2^m r_{12}^n
\end{eqnarray}
The nonlinear parameters \textit{i.e.} $\sigma$'s in eq. (11) and $\rho$'s in eq. (18) for $S$ and $P$ states respectively are optimized separately using Nelder-Mead algorithm \cite{nel36}. The linear variational parameters \textit{i.e.} $C_{lmn}$'s and $D_{lmn}$'s along with the energy eigenvalues are obtained by solving the generalized eigenvalue equation
\begin{eqnarray}
\underline{\underline{H}}~\underline{C}=E~\underline{\underline{S}}~\underline{C}
\end{eqnarray}
where $\underline{\underline{H}}$ is the Hamiltonian matrix, $\underline{\underline{S}}$ is the overlap matrix, $\underline{C}$ is the column matrix consisting of linear variational parameters and $E$ is the corresponding energy eigenvalue. The wave function is  normalized for each confining radius \textit{R} to account for the reorientation of charge distribution within the \textit{Wigner-Seitz} sphere. All computations are carried out in quadruple precision to ensure better numerical stability for extended Hylleraas basis sets within a finite domain.\\\\
The variational equation for n$l$ ($^2L$) states of one electron ion within ion-sphere of radius $R$ can be written as
\begin{eqnarray}\label{11}
\delta\int_{0}^{R} \left[\frac{1}{2}\left\{\left(\frac{\partial f}{\partial r}\right)^{2}+\frac{l(l+1)}{r^2}f^2\right\}+\left\lbrace V_{IS}(r)-E\right\rbrace f^{2}\right]r^{2}dr=0
\end{eqnarray}
where, the one-particle effective potential $V_{IS}(r)$ is taken from eq. (2) with $N_e$ = 1.
The radial function $f(r)$ is given by
\begin{eqnarray}
f(r)=(R-r)r^k\sum_{i}C_{i}e^{-\rho_{i}r}
\end{eqnarray}
where $k=0$ and 1 for $^2S$ and $^2P$ states respectively. In this calculation, we have taken 21 different nonlinear parameters ($\rho_{i}$'s) in a geometrical sequence $\rho_{i}=\rho_{i-1}\gamma$, $\gamma$ being the geometrical ratio \cite{pra37,jpb14}. Such choice of non-linear parameters enable us to cover the full region of space in a flexible manner by adjusting $\gamma$. The energy values $E$'s and linear co-efficients $C_{i}$'s are determined from eq. (19).\\\\
The truncation of wavefunction at a finite distance (eq. 3) imposes a thermodynamic pressure upon the ions which increases with increase of $n_e$ inside the sphere. We have calculated the pressure felt by all the hydrogen-like and helium-like ions in their respective ground state using the first law of thermodynamics. However,for excited states having finite lifetime, this approach is not valid as the equilibrium criteria is not maintained. Under an adiabatic approximation, the pressure on the ions in the ground state can be expressed as \cite{scr33}
\begin{eqnarray}
P=-\frac{1}{4 \pi R^2}\frac{dE}{dR}
\end{eqnarray}
\subsubsection*{Evaluation of two-electron integrals}
The correlated two-electron basis integrals arising in the present calculations are of the form
\begin{eqnarray*}
A(a,b,c;\alpha,\beta;R) = \int_0^R r_1^a~e^{-\alpha r_1}\int_0^R r_2^b~e^{-\beta r_2}\int_{|r_1-r_2|}^{r_1+r_2} r_{12}^c~dr_1 dr_2 dr_{12}
\end{eqnarray*}
\begin{equation}
=\int_0^R r_1^a~e^{-\alpha r_1}\int_0^{r_1} r_2^b~e^{-\beta r_2}\int_{r_1-r_2}^{r_1+r_2} r_{12}^c~dr_1 dr_2 dr_{12}+\int_0^R r_2^b~e^{-\beta r_2}\int_0^{r_2}r_1^a~e^{-\alpha r_1} \int_{r_2-r_1}^{r_1+r_2} r_{12}^c~dr_1 dr_2 dr_{12}
\end{equation}
For $S$ states, $a\geq 0, b\geq 0, c\geq 0$ while for higher angular momentum states (\textit{P, D etc.}), integrals with $a=-1$ also arises. After integration, the $r_{12}$ part of eq. (23) can be expanded as
\begin{eqnarray}
\frac{1}{n+1}\left[\left(r_1+r_1\right)^{n+1}-\left(r_1-r_1\right)^{n+1}\right]=\sum_{i=0}^{\frac{n}{2}}\frac{2.n!}{(2i+1)!(n-2i)!}r_1^{n-2i}r_2^{2i+1}~~~~~~[n~ even]
\end{eqnarray} 
For odd `$n$', the upper limit of the sum in the right hand side would be replaced by $\frac{n-1}{2}$. The integrals from eq. (23) then reduce to the form
\begin{eqnarray}
\int_0^y x^k~e^{-\lambda x}dx=\int_0^\infty x^k~e^{-\lambda x}dx-\int_y^\infty x^k~e^{-\lambda x}dx=\frac{k!}{\lambda^{k+1}}\left[1-e^{-\lambda y}\sum_{j=0}^{k}\frac{y^j\lambda ^j}{j!}\right]
\end{eqnarray}   
$\lambda$ is a positive real number and $k$ is a non-negative integer and we have used the standard integral
\begin{eqnarray}
\int_0^{\infty} x^k e^{-\lambda x}dx=\frac{k!}{\lambda ^{k+1}}
\end{eqnarray}
The integral $A(a,b,c;\alpha ,\beta; R)$ is now evaluated for two different cases.
\subsection*{Case I: $a\geq 0$, $b\geq 0$, $c\geq 0$}  
An exact analytical expression for $A(a,b,c;\alpha,\beta;R)$ corresponding to $a\geq 0, b\geq 0, c\geq 0$ has been derived in a straightforward way using eq. (25) and the numerical values are displayed in table-1. In the first column of table-1, different powers of $r_1$, $r_2$ and $r_{12}$ \textit{i.e.} $a$, $b$ and $c$ are given. For each set of ($a, b, c$), the non-linear parameters ($\alpha,\beta$) given in the second column of table-1 are varied from very low to high values as obtained from the optimized values corresponding to different cases in the present work. The values of $R$ varied in a wide range for each set of ($a, b, c$) and ($\alpha, \beta$) are given in the fourth column of table-1. All the values of integrals are given in the in the last column of table-1. We have checked the results with standard mathematical software (\textit{e.g. Maple}) to ensure the numerical stability of the expression for $A(a,b,c;\alpha,\beta;R)$ over the complete range of $R$ .   
\subsection*{Case II: $a=-1$, $b\geq 0$, $c\geq 0$} 
After full expansion of the integral $A(-1,b,c;\alpha,\beta;R)$ over $r_{12}$ and $r_{2}$ by using eqs. (24) and (25) an integral $I(\alpha,\beta;R)$ arise which takes the form 
\begin{eqnarray}
I(\alpha,\beta;R)=\int_0^R \frac{e^{-\alpha r_1}-e^{-(\alpha+\beta) r_1}}{r_1}dr_1
\end{eqnarray}
The above integral $I(\alpha,\beta;R)$ is actually a converging infinite series with oscillatory terms. We have tested the evaluation of the term $I(\alpha,\beta;R)$ in two different approaches.\\\\
\textit{i)} We can expand the exponential functions to evaluate the integral as
\begin{eqnarray}
\int_0^R \frac{e^{-\alpha r_1}-e^{-(\alpha+\beta) r_1}}{r_1}dr_1&=&\sum_{q=0}^{\infty}\int_0^R \frac{1}{r_1}\left[\frac{(-1)^q}{q!}\{\alpha ^q - (\alpha+\beta)^q \} r_1^q\right]dr_1  \nonumber \\
&=& \sum_{q=1}^{\infty}\frac{(-1)^q R^q}{q.q!}\left[\alpha ^q - (\alpha+\beta)^q \right] 
\end{eqnarray}
The expression (28) gives accurate value of integrals where the upper limit $R$ is small, but fails to produce results when $R$ is sufficiently high. \\
\textit{ii)} Alternatively, the integral $I(\alpha,\beta;R)$ may be written as
\begin{eqnarray}
\int_0^R \frac{e^{-\alpha r_1}-e^{-(\alpha+\beta) r_1}}{r_1}dr_1=\int_0^R \frac{e^{-\alpha r_1}}{r_1}\left(1-e^{-\beta r_1}\right)dr_1  = \sum_{q=1}^{\infty}\frac{(-1)^{q-1}\beta^q}{q!}\int_0^R r_1^{q-1} e^{-\alpha r_1}dr_1 
\end{eqnarray} 
The $r_1$-integral in the $r.h.s$ of equation (29) is then evaluated using expression (25).\\
The integral $I(\alpha,\beta;R)$ is calculated by using both the expressions given in eq. (28) and eq. (29). All the results corresponding to different sets of $(\alpha,\beta;R)$ are given in table 2 which shows excellent agreement among the results except for some high values of R used in eq. (28). On the other hand, eq. (29) yields excellent results irrespective of the values of the parameters $(\alpha,\beta)$ over the complete range of $R$. In eq. (28), a term $R^q$ appears in the numerator that increases with increase in $q$. For low values of $R$, this term is balanced by $q!$ in the denominator but for high $R$, a numerical instability appears because within the first few terms, $R^q$ bounces more rapidly than $q!$. In contrast, a term $\frac{R^j}{j!}e^{-\alpha R}$ appears in eq. (29) [after expanding the $r_1$-integral according to eq. (25)] which falls rapidly as $q$ increases due to the presence of the exponential term. To have a better understanding of the integrals, we have also checked the convergence of $I(\alpha,\beta;R)$ evaluated using eqs. (28) and (29) by increasing the number of terms in the infinite series and displayed the convergence behaviour in table 3 for $R=100.0$ and $0.2$ and two sets of $(\alpha,\beta)$. It appears from table 3 that for $R=100.0$ the values derived from eq. (28) are clearly not acceptable but for low $R$, the final results match exactly though the convergence is slow for equation (28). We have finally used eq. (29) to calculate the energy eigenvalues in the present work and taken 1000 terms in the corresponding infinite series to ensure the desired level of accuracy. In table 4 we have given the values of integral $A(-1,b,c;\alpha,\beta;R)$ corresponding to different sets of parameters. We have further observed that the integrals [eq. (29)] corresponding to $ R=100 $ yield same result as obtained by using eq. (26) for $ R=\infty. $ This is mention further that all the integrals are checked with standard mathematical software.\\
\section{Results and discussions}
The energy eigenvalues of He-like $C$, $Al$ and $Ar$ in $1sns~(^1S^e)~ [n=1-3]$ and $1sn'p~(^1P^o)~ [n'=2-4]$ states have been calculated within SCP environment using the IS potential. We have studied the convergence of the energy values \textit{w.r.t.} the number of terms (\textit{N}) in the wave function. Table 5 shows the convergence behaviour of $C^{4+}$ in $1s^2~(^1S^e)$ state for some selected values of $R$. We have obtained a similar convergence pattern for all the other ions and also for the excited states under consideration. The size of the basis has been extended systematically to $N = 161$ and 149 for $^1S^e$ and $^1P^o$ states respectively with $l+m+n=10$ [eqs. (11) and (18)]. The convergence of the energy values are obtained at least up to the sixth significant digits. In fact, for some cases \textit{e.g.} $1s^2~(^1S^e)$ state of $C^{4+}$ with $R=0.47$ a.u., we have obtained convergence of energy values up to the eighth decimal place, as is evident from table 5. The above observation ensures that the present method can deal with extended basis sets to yield sufficiently accurate energy values within a finite limit.\\\\
The energy values of He-like $C$, $Al$ and $Ar$ in $1sns~(^1S^e)~ [n=1-3]$ and  $1sn'p~(^1P^o)~ [n'=2-4]$ states within ion-sphere of different radii ($R$) are displayed in tables 6, 7 and 8 respectively. We have also listed the energies of respective H-like ions in $ns$ ($^{2}S$) [$n=1-2$] and $n'p$ ($^{2}P$) [$n'=2-3$] states. It is worthwhile to mention that under one-component plasma approximation, the IS radius for a two-electron ion would differ from that for a one-electron ion corresponding to the same plasma electron density. We see that as $n_e$  increases, the energy levels move towards continuum which is a clear manifestation of the positive nature of IS potential. To check the overall behavior of the results, we have plotted the energy values ($-E$) of bound $1sns$ ($^1S^e$) [$n=1-3$] and $1sn'p$ ($^1P^o$) [$n'=2-4$] states of $C^{4+}$ with respect to IS radius ($R$) in figure 1. It is evident from figure 1 that the energy values remain almost unaltered for a range of \textit{R} and after that rapidly approaches towards the destabilization limit. Hence the variation produces a `\textit{knee}' around some particular value of \textit{R}. For higher excited states, this `\textit{knee}' appears at a higher value of \textit{R}. All other ions also show same features. Similar behaviour of energy values of He-like ions inside a spherical impenetrable box (referred to as `Coulombic sphere' hereinafter) was reported in a recent publication \cite{scr33} where the potential inside the box was purely Coulombic. Within the ion-sphere, energy value of the positively charged ion is modified for two factors: 
\begin{enumerate}
\item The environment envisaged by IS potential which is governed by plasma electrn density \textit{and}
\item The truncation of wave function at a finite distance that generates a pressure on the system.
\end{enumerate} 
In order to asses the effect each factor on the energy eigenvalues, we have also studied separately the modification of energy values of two-electron ions due to the truncation of the wave function at different radii of Coulombic sphere. The ground state energy of a `\textit{free}' $C^{4+}$ ion where the wave function is infinitely extended is $-32.406247$ a.u. whereas within a Coulombic sphere and ion-sphere both having a radius of 20.0 a.u, the energy values are $-32.406247$ and $-31.806294$ a.u. respectively. It shows that for a large box radius, almost 100\% of the shift in the energy is due to the effect of plasma. The truncation of wave function becomes significant when the size of the sphere is reduced. At a radius of 0.7 a.u., the ground state energy values of $C^{4+}$ ion within Coulombic sphere and ion-sphere are $-31.192275$ and $-14.880628$ a.u. respectively which shows an effect of almost 7\% on the shift of energy level is coming from the truncation of wave function. This effect increases to 22.4\% and 26.2\% for truncation radius of 0.5 and 0.4692 a.u. respectively. \\
A closer look at the results quoted in tables $6-8$ leads us further to the following observations.
\begin{enumerate}
\item \textbf{Decrease in number of excited states}: For two-electron ions $C^{4+}, ~ Al^{11+}$ and $Ar^{16+}$ we see that as ${n_e}$ increases, the ions become less bound and also the number of excited states decreases. For example, $C^{4+}$ exists in the ground state up to $R=0.4692$ a.u. but $1s2s~(^1S^e)$ state ceases to exist after $R=0.9017$ a.u. and  $1s3s~(^1S^e)$ destabilizes after $R=1.3761$ a.u. Similar feature is observed for all the ions and also for $^1P^o$ states. For H-like ions of \textit{C, Al} and \textit{Ar}, the $2s$ state destabilizes much before $1s$ with increase of $R$.
\item \textbf{Reduction of ionization potential}: Ionization potential for a two-electron ion is defined as the amount of energy required to ionize one electron from the ground state ($ 1s^2 $). It is observed from tables $6-8$ for all the ions that with increase in plasma density, IP decreases and beyond certain density, the two-electron energy levels move above the one-electron threshold. We have studied the variation of IPD of two-electron ions \textit{w.r.t.} $n_e$ from the difference of IP within and without (\textit{i.e.} free case) the surrounding plasma environment. In figure-2, we have plotted the IP and IPD for $Al^{11+}$ as a function of plasma electron density. The energy required to ionize the outer electron from $1s3p(^1P^o) $ state \textit{i.e.} IP for $1s3p(^1P^o) $ state of $Al^{11+}$ and the corresponding IPD \textit{w.r.t.} $n_e$ are also included in figure-2. The effect of surrounding plasma on different two-electron energy levels should be different and consequently, IPDs should differ from one configuration to another. It is evident from figure-2 that the present observation corroborates this fact. The two-electron $^1P^o$ states would give rise to spectral lines via dipole transition until they merge into the one-electron continuum. For example, table-7 shows that $1s3p(^1P^o) $ state of $Al^{11+}$ can survive up to the density of $ 8.11 \times 10^{25} $/c.c. whereas it crosses the corresponding  $ 1s $ threshold after the density of $ 2.11 \times 10^{24} $/c.c and consequently, the $He_\beta$ line originating from $ 1s3p(^1P^o) \rightarrow 1s^2 (^1S^e) $ emission is not expected to be observed after this density. In table 9, we have listed the critical electron densities after which different spectral lines of H-like and He-like \textit{Al} disappear. The densities are calculated from IS radius according to both SP model and EK model of determining IPDs following eq. (4) and (5) respectively. For $Ly_\beta$ line of $ Al^{12+} $ and  $He_\beta$ line of $ Al^{11+} $, the present electron densities calculated by using SP model are in good agreement with experimental observation \cite{hoa}. For disappearance of $He_\gamma$ line, the only theoretical calculation of plasma density is due to Preston \textit{et al.} \cite{pre} where a possible range of densities is given. No experimental result is available for comparison in this context. Our results obtained by using SP model of IPD indicate that the $He_\gamma$ line of $ Al^{11+} $ would disappear after a plasma density of $ 5.0 \times 10^{23} $/c.c, as is given in table-9. We mention that the disappearance of both $Ly_\beta$ and $He_\beta$ lines are experimentally observed at the density of $ 2.2 \times 10^{24} $/c.c. whereas the $Ly_\beta$ line should survive more than the $He_\beta$ line. Present results along with ref. \cite{son} as depicted in table-9 establish the fact explicitly. A more accurate experimental measurement is therefore necessary for proper plasma diagnostics.\\ 
In an earlier experiment, Nantel \textit{et al.} \cite{nan23} observed $He_\alpha,He_\beta$ and $ He_\gamma $ lines of $C^{4+}$ at plasma density $1.5 \times 10^{21}$/c.c. In this experiment the densities corresponding to disappearance of such He-like lines are not explored. However, table-9 shows that He-like lines of $C^{4+}$ vanish well above the density of $1.5 \times 10^{21}$/c.c. Hence, the existence of such He-like lines of $C^{4+}$ at the density $1.5 \times 10^{21}$/c.c. as observed by Nantel \textit{et al.} \cite{nan23} are consistent with present calculations. Similar comparisons have been done with other earlier experiments of Saemann \textit{et al.} \cite{sae24} and Woolsey \textit{et al.} \cite{woo26} for spectral lines of $Al^{11+}$ and $Ar^{16+}$ respectively and the present results are in agreement with the experiments. Accurate measurement like the Orion laser experiment \cite{hoa} is necessary to confirm the present theoretical predictions for disappearance of spectral lines of $C^{4+}$ and $Ar^{16+}$.
\item \textbf{Quasi-bound states of two-electron ions}: Quasi-bound states or continuum bound states may be found in continuous part of the spectra for electronic confinement under different potentials \cite{del} and has also been observed experimentally \cite{cap}. These states have great structural similarity with the discrete energy levels. For a two-electron ion, the ground state and all singly excited energy levels, in general, lie below the first ionization threshold. Tables 6-8 show that for high values `$R$' (\textit{i.e.} almost free case), this feature is  maintained for all the ions but as $R$ decreases, all singly excited states of two-electron ions become less bound more rapidly than the respective one-electron ion. For example, at $R=20.0$ a.u. the energy values of $C^{4+}$ as reported in table-6 lie below the $1s$ threshold of $C^{5+}$. At $R = 5.0$ a.u. the $1s4p~(^1P^o)$ state moves above the $1s$ threshold but lies below 2\textit{s} threshold. Similarly, at $R = 2.0$ a.u. $1s3s~(^1S^e)$ and $1s3p~(^ 1P^o)$ states lie above the $ 1s $ threshold and below $2s$ threshold. At $R<1.7$ a.u. the $2s$ level of $C^{5+} $ destabilizes and we observe well-converged (up to 7th significant digits) energy level of $1s4p~(^1P^o)$ state of $C^{4+}$ embedded in one-electron continuum. Similar feature is obtained for other ions also and is being reported for the first time in SCP.
\item \textbf{Incidental degeneracy and level crossing}: For a free two-electron ion, the energy value of $1s2s~(^1S^e)$ state is more negative than $1s2p~(^1P^o)$ state. Tables 6-8 establish this fact for high values of $R$ corresponding to all the two-electron ions. For example, the $1s2s~(^1S^e)$ level of $C^{4+}$ lies below $1s2p~(^1P^o)$ level for IS radius up to $R=2.0$ a.u. At $R=1.5$ a.u., the $1s2s~(^1S^e)$ state moves above the $1s2p~(^1P^o)$ level. These results show that an `$incidental~ degeneracy$' \cite{sen38} has taken place for $1s2s~(^1S^e)$ and $1s2p~(^1P^o)$ states of $C^{4+}$ at some value of $R$ between 1.5 and 2.0. a.u. and then a `$level~ crossing$' occurs between two states having different symmetry properties. The phenomenon of $incidental~degeneracy$ was reported in case of shell-confined hydrogen atom by Sen \cite{sen38} where two initially non-degenerate states are brought to a same energy level by adjusting external parameters. For a two-electron ion, we report \textit{incidental degeneracy} for the first time within SCP environment. After the \textit{level crossing}, the $1s2s~(^1S^e)$ state of $C^{4+}$ destabilizes (at $R=0.9017$ a.u.) much before $1s2p~(^1P^o)$ state (at $R=0.797$ a.u.). Similarly, for $1s3s~(^1S^e)$ and $1s3p~(^1P^o)$ states of $C^{4+}$, $incidental~ degeneracy$ and subsequent $level~ crossing$ are observed at a value of $R$ lying somewhere between 5.0 and 3.0 a.u. We observe similar phenomena for other ions also. For $2s$ and $ 2p $ states of H-like ions embedded in SCP, `$incidental~ degeneracy$' and `$level~ crossing$' phenomena are evident from tables $6-8$ and being reported here for the first time. 
\end{enumerate}
In table 10, we have displayed the comparison of present results for transition energies of $C^{4+},~Al^{11+}$ and $Ar^{16+}$ with experimental measurements \cite{nan23,sae24,woo26} and existing theoretical calculations \cite{pkm27}. Table-10 clearly depicts that the present results are in good agreement with the experimental measurements as compared to other theoretical results. The difference between theoretical and experimental results is due to the relativistic effects which is more significant for $Al^{11+}$ and $Ar^{16+}$ than $C^{4+}$. For $Ar^{16+}$ confined in ion-sphere, Sil \textit{et al.} \cite{pkm27} have made a comparison of theoretically calculated transition energies with the experiments but the plasma electron density reported by them is one order less than the experimental condition of $1.0 \times 10^{24}$ per c.c. \cite{woo26} and the IS radius of 2.057 a.u. \cite{pkm27} also does not correspond to any of the densities. Actually, according to eq. (4), IS radius of 2.057 a.u. corresponds to a plasma electron density of $2.96 \times 10^{24}$ per c.c. for $Ar^{16+}$. We have calculated the energies corresponding to $1s^2 \rightarrow 1s$n\textit{p} [n $=2-4$] transitions of $Ar^{16+}$ for both experimental \cite{woo26} and theoretical \cite{pkm27} conditions and given the values in table-10. It is seen from table-10 that the $1s^2 \rightarrow 1s5p$ transition energies estimated from present results are more away from experiments as compared to other transitions. We note that present size of the trial wave function for a high-lying state like $1s5p~(^1P^o)$ is not sufficient and needs to be increased to obtain greater accuracy.\\\\
The thermodynamic pressure experienced by H-like and He-like $C,~Al$ and $Ar$ in their respective ground states have been calculated for different values of IS radius $R$ using eq. (22) and the results are given in table-11. It is clear from table-11 that as $n_e$ increases, the pressure upon the ion increases and the ion moves towards destabilization. We observe that for a low value of $n_e$, the pressure upon the one-electron ion is higher than the respective two-electron ion and after a certain increase of $n_e$, the pressure on the two-electron ion exceeds the pressure experienced by the corresponding one-electron ion. With a view to studying the variation of thermodynamic pressure ($P$) with respect to the IS volume ($V$) under an adiabatic expansion, we have tried to fit the results for the two-electron ions obtained from the present calculations according to the ideal gas relation,
\begin{eqnarray}
PV^\gamma = constant ~~~~~~~~~ or~~~~~~~~~  ln P + \gamma ln V = constant
\end{eqnarray} 
where $\gamma$ is the ratio of two specific heats. From a least square fit of $ln~P~~ vs. ~~ln~V$ plot, the value of $\gamma$ comes out to be close to 1.4 for all the two-electron ions. To be precise, for $C^{4+},~Al^{11+}$ and $Ar^{16+}$, the values of $\gamma$ are 1.41, 1.37 and 1.37 respectively.  

\section{Conclusion}
Accurate analytical evaluation of the two-electron correlated integrals in Hyllerass coordinates within a finite limit has been performed. The intricacies of such calculations have been discussed in detail and the general applicability of these integrals has been established for arbitrary values of physically acceptable parameters. This methodology has immense potential to be useful for evaluation of the energy values and other spectral properties for three-body ionic and exotic systems placed within different external confinements such as strongly and moderately coupled plasma, fullerene cages, barrier potential, potential well \textit{etc.} With the recent advancement in experimental technique, the present methodology becomes relevant for calculating accurate plasma electron density from the spectral analysis of hydrogen and helium-like ions. We conclude that the ion-sphere potential where the electron density is calculated by using the SP model of IPDs provides a realistic picture of ions embedded in SCP environment. The present non-relativistic results reported here can be useful for plasma diagnostics and the non-relativistic energy values can serve as benchmark for future calculations to estimate relativistic and QED effects on two-electron ions within finite domain.\\\\
\textbf{{Acknowledgment}}\\\\
SB acknowledges financial support under grant number PSW-160/14-15(ERO) from University Grants Commission, Govt. of India. TKM acknowledges financial support under grant number 37(3)/14/27/2014-BRNS from the Department of Atomic Energy, BRNS, Govt. of India.

\begin{table}[!th]
\begin{center}
\begin{footnotesize}
\caption {Values of integral $A(a,b,c;\alpha,\beta;R)$ with $a\geq 0, b\geq 0, c\geq 0$. The notation $x(y)$ indicates $x \times 10^{y}$.}
\begin{tabular}{c c c l l}\vspace{-0.1cm}\\
  \hline\hline\vspace{-0.3cm}\\
 $(a,b,c)$ & $\alpha$&$\beta$ & $R$ & $A(a,b,c;\alpha,\beta;R)$\\
 \hline\vspace{-0.3cm}\\
(0,0,0) & 0.62450527& 0.41287135             & 100.0~             & 0.7477~2634~8987~8847 ~(+01)\\
  &                  && 2.0~                  & 0.1578~7453~6391~8980 ~(+01) \\
 &  &                & 0.2                   & 0.4688~3009~2327~4260 ~($-$02) \\ 
 & 8.92934001& 5.97270373                 & 100.0       & 0.2516~4820~0437~6827~~($-$02) \\
 &&                  & 2.0                   & 0.2516~4546~4241~1431~~($-$02)\\
 & &                 & 0.2                   & 0.9770~3676~6565~7275~~($-$03)\\
 & 17.42010556& 10.32300145  & 100.0                 & 0.4008~8345~0596~5748~~($-$03)\\
 &&                  & 2.0~                  & 0.4008~8344~9906~8468 ~($-$03)\\
 &&                  & 0.2~                  & 0.3006~0461~4922~4955~~($-$03)\\
 (2,3,1) & 0.62450527 & 0.41287135     & 100.0~                & 0.1578~2446~3158~5587 ~(+06)\\
 &&                  & 2.0~                  & 0.9755~4412~5697~4906~~(+01)\\
 &&                  & 0.2                   & 0.4326~2415~0573~9082~~($-$07) \\
 & 8.92934001& 5.97270373  & 100.0                 & 0.5960~1910~9088~6659~~($-$05) \\
  &&                  & 2.0~                  & 0.5913~0152~3072~3085~~($-$05)\\
 & &                 & 0.2                   & 0.4824~7810~4677~3954~~($-$08) \\
  & 17.42010556& 10.32300145    & 100.0                 & 0.2667~8377~6884~8479~~($-$07)\\ 
 &&                  & 2.0~                  & 0.2667~8110~4519~4153~~($-$07)\\
 & &                 & 0.2~                  & 0.7198~6505~5902~0197~~($-$09)\\
 (3,4,6) & 0.62450527& 0.41287135   & 100.0~                & 0.5959~0474~3300~4562 ~(+14)\\
 &&                  & 2.0~                  & 0.3133~2861~0504~5369 ~(+04)\\
 &&                  & 0.2                   & 0.1612~2935~4296~9925~~($-$11)\\
  & 8.92934001& 5.97270373  & 100.0                 & 0.1741~4651~3311~5703~~($-$04) \\
 &&                  & 2.0~                  & 0.1311~5563~4874~3337 ~($-$04)\\
 &&                  & 0.2                   & 0.1427~2637~0766~0009~~($-$12) \\
  & 17.42010556& 10.32300145  & 100.0                 & 0.1312~3966~9657~3856~~($-$08)\\
 &&                  & 2.0~                  & 0.1305~9386~1401~5103~~($-$08)\\
 &&                 & 0.2                    & 0.1601~5416~4752~0373~~($-$13) \\
\hline\hline
\end{tabular}
\end{footnotesize}
\end{center}
\end{table}

\begin{table}[!th]
\begin{center}
\begin{footnotesize}
\caption {\rm {Values of integral $I(\alpha,\beta;R)$. Results obtained by using eq. (28) and eq. (29) are given in consecutive rows respectively. The notation $x(y)$ indicates $x \times 10^{y}$.}}
\begin{tabular}{c c l l}\\
  \hline\hline\vspace{-.3cm}\\
 $\alpha$ & $\beta$ & $R$ &$I(\alpha,\beta;R)$\\
 \hline\vspace{-.3cm}\\
  0.62450527  &  0.41287135 &100.0     & 0.2910~1096~5167~7626~(+9) \\
  &&                                   & 0.5074~9055~6270~2974 \\
  && 2.0                               & 0.4049~6455~3461~7721 \\
  &&                                   & 0.4049~6455~3461~7721 \\
  && 0.2                               & 0.0760~8455~8526~4357 \\
  &&                                   & 0.0760~8455~8526~4357 \\
  8.92934001  & 5.97270373 & 100.0     & 0.2617~9165~1879~3351~(+602) \\
  &&                                   & 0.5121~5588~2201~4390 \\
  && 2.0                               & 0.5121~5588~1268~8812 \\
  &&                                   & 0.5121~5588~1268~8812 \\
  && 0.2                               & 0.4595~0838~7239~3843 \\
  &&                                   & 0.4595~0838~7239~3843 \\
  17.42010556   & 10.32300145  &100.0  & 0.9419~3093~2537~1151~(+872) \\
  &&                                   & 0.4653~6238~2193~5131~ \\
  && 2.0                               & 0.4653~6238~2193~5052  \\
  &&                                   & 0.4653~6238~2193~5131 \\
  && 0.2                               & 0.4588~5876~8812~0597  \\
  &&                                   & 0.4588~5876~8812~0597 \\
 \hline\hline
\end{tabular}
\end{footnotesize}
\end{center}
\end{table}

\begin{table}[!th]
\begin{center}
\begin{footnotesize}
\caption {\rm {Convergence of the integral $I(\alpha,\beta;R)$ \textit{w.r.t.} the number of terms (\textit{q}) in the infinite series using equations (28) and (29). The notation $x(y)$ indicates $x \times 10^{y}$.}}
\begin{tabular}{c c l l l c}\\
  \hline\hline\vspace{-.3cm}\\
  &  &  &  &\multicolumn{2}{c}{$I(\alpha,\beta;R)$}\\\cline{5-6}\vspace{-.2cm}\\
 $\alpha$ & $\beta$ & $R$ & \textit{q} & \multicolumn{1}{c}{Eq. (28)} & \multicolumn{1}{c}{Eq. (29)}\\
 \hline\vspace{-.3cm}\\
 0.62450527  & 0.41287135 & 100 & 10   & -0.3568~5265~8756~6828~(+13) & 0.5068~9300~3143~0272 \\
 &                        &     & 20   & -0.3554~0371~2221~5343~(+21) & 0.5074~8564~3805~4664 \\
 &                        &     & 50   & -0.2755~0482~1577~0421~(+35) & 0.5074~9055~6262~6217 \\
 &                        &     & 100  & -0.2124~7555~4344~8736~(+42) & 0.5074~9055~6270~2974 \\
 &                        &     & 1000 &  0.2910~1096~5167~7626~~(+09) & 0.5074~9055~6270~2974 \\
 &                        & 0.2 & 10   &  0.0760~8455~8526~4356 & 0.0760~8455~8526~4357 \\
 &                        &     & 20   &  0.0760~8455~8526~4357 & 0.0760~8455~8526~4357 \\
 &                        &     & 50   &  0.0760~8455~8526~4357 & 0.0760~8455~8526~4357 \\
 &                        &     & 100  &  0.0760~8455~8526~4357 & 0.0760~8455~8526~4357 \\
 &                        &     & 1000 &  0.0760~8455~8526~4357 & 0.0760~8455~8526~4357 \\
17.42010556 & 10.32300145 & 100 & 10   & -0.7343~5142~7504~1944~(+27) & 0.4651~7579~3447~4452 \\
 &                        &     & 20   & -0.1488~0713~2757~9564~(+50) & 0.4653~6186~8073~0945 \\
 &                        &     & 50   & -0.9285~2359~3991~9110~(+106) & 0.4653~6238~2193~4812 \\
 &                        &     & 100  & -0.2137~8120~9553~6411~(+185) & 0.4653~6238~2193~5131 \\
 &                        &     & 1000 & -0.2609~9042~7983~5017~(+873) & 0.4653~6238~2193~5131 \\
 &                        & 0.2 & 10   & 0.2166~5736~2229~1603 & 0.4588~5852~3846~0869 \\
 &                        &     & 20   & 0.4588~5558~5227~7264 & 0.4588~5876~8812~0596 \\
 &                        &     & 50   & 0.4588~5876~8812~0597 & 0.4588~5876~8812~0597 \\
 &                        &     & 100  & 0.4588~5876~8812~0597 & 0.4588~5876~8812~0597 \\
 &                        &     & 1000 & 0.4588~5876~8812~0597 & 0.4588~5876~8812~0597 \\
 \hline\hline
\end{tabular}
\end{footnotesize}
\end{center}
\end{table}

\begin{table}[!th]
\begin{center}
\begin{footnotesize}
\caption {\rm {Values of integral $A(a,b,c;\alpha,\beta;R)$ with $a = -1, b\geq 0, c\geq 0$. The notation $x(y)$ indicates $x \times 10^{y}$.}}
\begin{tabular}{c c c l l}\\
  \hline\hline\vspace{-0.3cm}\\
 $(a,b,c)$ & $\alpha$&$\beta$ & $R$ & $A(-1,b,c;\alpha,\beta;R)$\\
 \hline\vspace{-0.3cm}\\
 (-1,0,0) & 0.62450527& 0.41287135    & 100.0~                & 0.5954~2692~0365~7702~(+01)\\
 &                  && 2.0~                   & 0.2328~7535~9564~1012~(+01)\\
 &  &                & 0.2                    & 0.5410~3409~5795~4025~($-$01)\\
  & 8.92934001& 5.97270373    & 100.0                  & 0.2871~3769~8867~2015~($-$01) \\
 &&                  & 2.0                    & 0.2871~3526~4943~5828~($-$01)\\
 & &                 & 0.2                    & 0.1630~9271~2387~9263~($-$01)\\
 & 17.42010556& 10.32300145  & 100.0                  & 0.8733~9216~7581~8471~($-$02)\\
 &&                  & 2.0~                   & 0.8733~9216~6380~3946~($-$02)\\
 &&                  & 0.2~                   & 0.7244~1513~1903~8541~($-$02)\\
 (-1,3,0) & 0.62450527 & 0.41287135  & 100.0~ & 0.6567~2317~0086~4201~(+03)\\
 &&                  & 2.0~                   & 0.4615~4100~9066~9881~(+01)\\
 &&                  & 0.2                    & 0.1352~9814~0488~3989~($-$03)\\ 
& 8.92934001& 5.97270373   & 100.0                  & 0.1048~5277~9488~0441~($-$02)\\
 &&                  & 2.0~                   & 0.1046~0034~4655~7635~($-$02)\\
 & &                 & 0.2                    & 0.2845~3033~4471~0229~($-$04)\\
 & 17.42010556& 10.32300145 &     100.0                  & 0.6034~8938~8061~1726~($-$04)\\  
 &&                  & 2.0~                   & 0.6034~8827~2996~3596~($-$04)\\
 & &                 & 0.2~                   & 0.8940~2524~7131~8926~($-$05)\\
(-1,6,4) & 0.62450527& 0.41287135 &  100.0~                 & 0.1994~0115~8159~1710~(+12)\\
 &&                  & 2.0~                   & 0.3251~2782~4902~2441~(+03)\\
 &&                  & 0.2                    & 0.1201~4454~0849~6434~($-$08)\\
  & 8.92934001& 5.97270373 &      100.0     & 0.2402~8418~0970~4088~($-$02)\\
 &&                  & 2.0~                   & 0.1563~9297~5468~9225~($-$02)\\
 &&                  & 0.2                    & 0.1788~1043~5252~0599~($-$09)\\
 & 17.42010556& 10.32300145 &  100.0                  & 0.2982~9940~1992~8102~($-$05)\\
 &&                  & 2.0~                   & 0.2960~5934~0875~9482~($-$05)\\
 &&                 & 0.2                     & 0.4140~7243~9026~4580~($-$10) \\
\hline\hline
 \end{tabular}
 \end{footnotesize}
\end{center}
\end{table}
\begin{table}[!th]
\begin{center}
\caption {\rm {Convergence of energy values ($-E$ a.u.) of $1s^2(^1S^e)$ state of $C^{4+}$ with respect to number of terms (\textit{N}) in wave function within Ion-sphere radius \textit{R} a.u.}}
\begin{footnotesize}
\begin{tabular}{c l c c c c c}\\
  \hline\hline\vspace{-0.2cm}\\
 & & \multicolumn{5}{c}{$-E$ for two-electron ions}\\
  \cline{3-7}\vspace{-0.2cm}\\
   State &\textit{N} &$R = 20.0$     & 0.7        & 0.5 & 0.47  & 0.4692 \\  
   \hline\vspace{-0.2cm}\\
  $1s^2(^1S^e)$&13  & 31.8060 7622 & 14.8798 5637 & 3.4091 6161 & 0.1035 1773  & 0.0053 4000 \\
               & 22 & 31.8062 6559 & 14.8805 9032 & 3.4092 5286 & 0.1037 0540  & 0.0055 3914 \\
               & 34 & 31.8062 9082 & 14.8806 1939 & 3.4092 6383 & 0.1037 1810  & 0.0055 5223 \\
               & 50 & 31.8062 9351 & 14.8806 2562 & 3.4092 6610 & 0.1037 1980  & 0.0055 5423 \\
               & 70 & 31.8062 9412 & 14.8806 2707 & 3.4092 6664 & 0.1037 2017  & 0.0055 5462 \\
               & 95 & 31.8062 9431 & 14.8806 2746 & 3.4092 6678 & 0.1037 2026  & 0.0055 5471\\
              & 125 & 31.8062 9439 & 14.8806 2759 & 3.4092 6682 & 0.1037 2029  & 0.0055 5474 \\
              & 161 & 31.8062 9443 & 14.8806 2763 & 3.4092 6683 & 0.1037 2029  & 0.0055 5475 \\
  \hline\hline
 \end{tabular}
\end{footnotesize}
\end{center}
\end{table}

\begin{sidewaystable}[tbp]
\begin{center}
\caption {\rm {Energy eigenvalues ($-E$ a.u.) of $ C^{4+} $ and $ C^{5+} $ within ion-sphere of Radius $R$ a.u. Densities (per c.c.) are determined from radius by using eq. (4). The uncertainty of the calculated energy values is of the order of $10^{-6}$ a.u.}}
\begin{scriptsize}
\begin{tabular}{l l c c c c c c c l c c c c}\\
  \hline\hline\vspace{-0.2cm}\\
 Plasma & \multicolumn{7}{c}{$-E~$ for two-electron ions}&&\multicolumn{5}{c}{$-E$ for one-electron ions}\\
  \cline{2-8}\cline{10-14}\vspace{-0.2cm}\\
 density& R & $1s^2$ & $1s2s$ & $1s3s$ & $1s2p$ & $1s3p$ & $1s4p$ && R & $E_{1s}$ & $E_{2s}$ &$ E_{2p}$& $E_{3p}$\\
 \hline\vspace{-0.2cm}\\
 &$C^{4+}$&&&&&&&&$C^{5+}$&&&\\
  8.05(20)  & 20.0  & 31.806 294 & 20.622 433 & 18.819 815 & 20.493 660 & 18.782 365 & 18.183 990& & 21.544 & 17.651 901 &4.152 167 & 4.151 006 & 1.298 528\\
 1.50(21)    &16.256 & 31.668 147 &             &             & 20.355 753 & 18.645 771 & 18.051 134 && 17.511 &17.571 741 & 4.072 241 & 4.070 983 & 1.216 051\\
 6.44(21)   & 10.0  & 31.206 629 & 20.025 353 & 18.234 022 & 19.895 956 & 18.195 421 & 17.626 905 && 10.772 &17.303 930 & 3.806 085 & 3.804 210 & 0.941 359\\
 1.88(22)  & 7.0   &             &             &             &             & 17.710 079 & 17.127 014 && 7.540 &17.005 859 & 3.512 120 & 3.508 790 & 0.638 130\\
 5.15(22)  & 5.0   &	30.009 311 & 18.848 891 &	17.115 098 & 18.714 486 &	17.073 177 & 16.164 573 &&5.386 &16.608 855 & 3.126 251 & 3.119 111 & 0.239 771\\
 8.32(22)&&&&&&&&&4.591&&&2.885 861&0.001 097\\
 1.01(23)   & 4.0   &&&&&                                                                 & 14.979 005 && 4.309 & 16.262 009 & 2.796 224 & 2.783 284 &\\
 2.39(23)   & 3.0   & 28.420 442 & 17.332 928 & 15.017 054 & 17.183 256 & 15.108 681 & 12.307 491 && 3.232 & 15.685 381 & 2.266 340 & 2.237 178\\
 4.12(23)   & 2.5   &             &             & 13.431 934 &             & 13.720 854 & 9.562 152 &&2.693  & 15.225 721 & 1.855 508 & 1.812 626\\
 6.05(23)   & 2.2   &&&&&                                                                 & 6.922 023 &&2.370 & 14.850 960 & 1.512 931 & 1.468 923\\
 8.05(23)   & 2.0   &             & 15.256 681 & 10.320 034 & 15.180 308 & 11.114 072 & 4.454 707 &&2.154 & 14.539 678 & 1.207 836 & 1.175 386\\
 1.10(24)   & 1.8   &&                          & 8.209 279  &&                          & 1.315 545 &&1.939 & 14.160 665 & 0.797 941 & 0.799 140\\
 1.31(24)     & 1.7   &&                          & 6.840 488  &             & 8.309 520  & 0.736 221 &&1.831 & 13.938 515 & 0.529 722 & 0.563 106\\
  1.57(24)   & 1.6   & 24.999 764 &&&                                       & 7.004 681  &&&1.723 & 13.689 353 & 0.200 592 & 0.283 113\\
  1.58(24)   & 1.596 &&&&&                                                         & 0.030 797 &&1.719 & 13.678 753 & 0.186 970 & 0.271 734\\
1.85(24)  &&&&&&&&&1.632&&&0.003 072&\\
  1.91(24)  & 1.5   &             & 12.478 314 & 3.173 818  &12.790 730 &&& &1.616 &13.407 990 & \\
  2.35(24)  & 1.4 &&                            & 0.691 166  &             & 3.506 751  &&&1.508 & 13.087 774&\\
  2.38(24)  &1.3937 &&                          & 0.516 288  &&&& &1.501 & 13.066 138 &\\
  2.47(24)  &1.3761 &&                          & 0.015 255 &&&& &1.482 &13.004 619 &\\
  2.93(24) & 1.3   &             & 10.292 255 &&                          & 1.124 349  && &1.400 &12.720 153 &\\
  3.20(24)   & 1.263 &&&&                                                 & 0.096 911 & & &1.360 &12.569 908 &\\
  3.73(24)  & 1.2   &             & 8.722 057  &             & 9.902 622  &&& &1.293 & 12.293 555 &\\
  6.44(24)  & 1.0   & 20.674 940  & 3.865 288 &              & 6.473 522  &&& &1.077 &11.190 638 &\\
  7.52(24)  & 0.95  &             & 2.091 381 &                               &   &&& &1.023 &10.841 350&\\
  8.55(24)  & 0.91  &             & 0.434 054 &&&&&& 0.980 & 10.532 355 &\\
  8.79(24)  & 0.9017&             & 0.059 416 &&&&& &0.971  & 10.464 484 &\\
  8.84(24)  & 0.9   &&&                                       & 3.864 058  &&&& 0.969 &10.450 413 &\\
  1.25(25)  & 0.8   &&&                                       & 0.209 360  &&&& 0.862 & 9.499 773 &\\
  1.27(25)  & 0.797  &&&                                      & 0.077 557  &&&& 0.858 & 9.466 811 &\\
  1.88(25)  & 0.7   & 14.880 628 &&&&&&& 0.754 & 8.198 121 &\\
  5.15(25)  & 0.5   & 3.409 267  &&&&&&& 0.539 & 2.956 871 &\\
  6.21(25)  & 0.47  &0.103 720 &&&&&&& 0.5063 & 1.497 759 &\\
  6.23(25)  &0.4695 &0.042 434 &&&&&&& 0.5057 & 1.470 782 &\\
  6.24(25)  &0.4692 &0.005 555 &&&&&&& 0.5054 & 1.454 571 &\\
                    \hline\hline
 \end{tabular}
\end{scriptsize}
\end{center}
\end{sidewaystable}

\begin{sidewaystable}[tbp]
\caption {\rm {Energy eigenvalues ($-E$ a.u.) of $ Al^{11+} $ and $ Al^{12+} $ within ion-sphere of Radius $R$ a.u. Densities (per c.c.) are determined from radius by using eq. (4). The uncertainty of the calculated energy values is of the order of $10^{-6}$ a.u.}}
\begin{center}
\begin{scriptsize}
\begin{tabular}{l l c c c c c c c l c c c c}\\
\hline\vspace{-0.2cm}\\
 Plasma & \multicolumn{7}{c}{$-E$ for two-electron ions}&&\multicolumn{5}{c}{$-E~$ for one-electron ions}\\
  \cline{2-8}\cline{10-14}\vspace{-0.2cm}\\
 density& R & $1s^2$ & $1s2s$ & $1s3s$ & $1s2p$ & $1s3p$ & $1s4p$ && R & $E_{1s}$ & $E_{2s}$ &$ E_{2p}$& $E_{3p}$\\
 \hline\vspace{-0.2cm}\\ 
         &$Al^{11+}$&&&&&&&&$Al^{12+}$&\\          
 2.21(21)  & 20.0 & 159.382 029 & 101.075 286 & 90.920 760 & 100.751 849 & 90.826 863 & 87.343 470 && 20.589  & 83.625 742 & 20.250 901 & 20.250 869 & 8.508 494\\
 1.77(22)  & 10.0 & 157.732 211 & 99.426 741 & 89.277 651  & 99.102 949  & 89.183 005 & 85.713 789 && 10.294 & 82.751 557 & 19.377 826 & 19.377 385 & 7.638 134  \\
 5.17(22)  & 7.0   &156.318 323 & 98.015 642 & 87.878 476 &              &             &           & & 7.206 & 82.002 366 & 18.631 067 & 18.629 929 & 6.898 570\\
 1.00(23)  & 5.617 &155.158 153 &             &           & 96.533 764  & 86.640 729 & 83.249 437 && 5.782 & 81.387 603 & 18.019 768 & 18.017 402 & 6.297 680 \\
 1.42(23)  & 5.0   &             &             &          & 95.811 753  & 85.932 485 & 82.582 105 && 5.147 & 81.003 693  & 17.638 853 & 17.635 633 & 5.926 068 \\
 5.00(23)   & 3.285 &150.992 205 &             &          & 92.391 567  & 82.632 916 & 79.441 867 && 3.382 & 79.179 940 & 15.840 819 & 15.830 292 & 4.207 565\\
 6.56(23)  & 3.0   &150.039 702 & 91.786 850 & 81.867 981 &              &             &       & &3.088  & 78.675 129 & 15.347 266 & 15.332 246 & 3.746 670 \\
 1.00(24)  & 2.607 &148.385 508 &             &                   & 89.814 745  & 80.216 318 & && 2.684 & 77.798 403 & 14.495 335 & 14.473 891 & 2.966 481\\ 
 2.21(24)  & 2.0   &144.557 994 & 86.435 075 & 76.751 134 & 86.060 502  & 76.632 056 & 70.991 150 && 2.058 & 75.769 505 & 12.554 606 & 12.501 889 & 1.209 389 \\
3.31(24) & &&&&&&&& 1.8 &&& 11.309 886 & 0.084 801\\
 5.25(24)  & 1.5   &139.093 636 & 81.216 562 & 70.476 891 & 80.776 714  & 70.672 695 & 59.958 894 && 1.544 & 72.871 972 & 9.877 422 & 9.760 672 \\
 1.77(25)  & 1.0   &128.240 365 & 70.674 599 & 51.222 379 & 70.269 488  & 53.774 530 & 27.642 101 && 1.029 & 67.112 431 & 4.647 358 & 4.445 998\\
 2.43(25)  & 0.9   &             &             &             & 66.602 776 &        & 13.910 306 && 0.926 & 65.206 039 & 2.699 333 & 2.596 821\\
 3.10(25)  & 0.83  &             &             & 34.498 851 & 63.390 834 &          & 1.358 983 && 0.854 & 63.604 555 & 0.880 652 & 0.952 002 \\
3.21(25)  & 0.82  &             &             &             &              &         & 0.585 110 & & 0.844  & 63.354 014 & 0.578 268 & 0.692 836\\
 3.31(25)        & 0.812 &             &             &             &              &     & 0.088 777 && 0.836 & 63.149 234 & 0.327 244 & 0.479 227 \\
 3.46(25)  & 0.8   &             & 61.584 852 & 30.304 265 &&&&& 0.823 & 62.834 616 & &0.011 954\\
 5.17(25)  & 0.7   &114.496 531  & 54.024 877 & 11.937 344 & 55.190 310   & 21.877 996&&& 0.721 & 59.806 678 &\\
 5.64(25)  & 0.68  &             &             & 7.201 566  & &&&& 0.700 & 59.097 854 &\\
 6.16(25)  & 0.66  &             &             & 2.007 102  & & &&& 0.6794 & 58.347 614 &\\
 6.18(25)  &0.6594 &             &             & 1.843 590  & & &&& 0.6788 & 58.324 454 &\\
 6.37(25)  &0.65283&&                          & 0.022 248 &&&&   & 0.672 & 58.067 871 &\\
 6.45(25)  & 0.65  &&&&                                                     & 11.929 838 &&& 0.669 & 57.955 830 &\\
 7.81(25)  & 0.61  &&&&                                                     & 2.170 876  && &0.628 &56.265 059 &\\
 8.08(25)  & 0.603 &&&&                                                     & 0.262 866  &&& 0.621 &55.946 962 &\\
 8.10(25)  & 0.6025&&&&                                                     & 0.124 062  &&& 0.6202 &55.923 954 &\\
 8.11(25)  & 0.6021&&&&                                                     & 0.012 689  &&& 0.6198 &55.905 567 &\\
 1.41(26)  & 0.5   & 96.523 118 & 22.189 554 &             & 30.281 348  &             &&& 0.515 &50.268 732 &\\
 2.08(26)  & 0.44  &             & 2.370 513 &&&&&& 0.453 &45.729 585 &\\
 2.14(26)  & 0.436 &             & 0.724 569 &&&&&& 0.449 &45.379 873 &\\
 2.15(26)  & 0.4351&             & 0.347 428 &&&&& &0.448 &45.300 161 &\\
 2.16(26)  &0.43434&             & 0.026 971 &&&&&& 0.447 &45.232 635 &\\             
2.77(26)   & 0.4   & &&                                     & 3.174 966 &&&& 0.412  & 41.877 044 & \\
2.87(26)   & 0.395 & &&                                     & 1.274 781 &&&& 0.407 & 41.331 971 & \\
2.95(26)   & 0.3918& &&                                     & 0.020 739 &&&& 0.403 & 40.974 430 &\\
6.56(26)   & 0.3   & 49.123 982 &&&&&&& 0.309 & 26.244 361 &\\
1.28(27)   & 0.24  & 8.502 491  &&&&&&& 0.247 & 6.577 159 &\\
1.419(27)  & 0.232 & 0.362 136  &&&&& && 0.239 & 2.673 160 &\\
1.424(27)  & 0.2317& 0.038 819  &&&&&&& 0.238 & 2.518 185 & \\
 \hline\hline
 \end{tabular}
\end{scriptsize}
\end{center}
\end{sidewaystable}

\begin{sidewaystable}[tbp]
\caption {\rm {Energy eigenvalues ($-E$ a.u.) of $ Ar^{16+} $ and $ Ar^{17+} $ within ion-sphere of Radius $R$ a.u. Densities (per c.c.) are determined from radius by using eq. (4). The uncertainty of the calculated energy values is of the order of $10^{-6}$ a.u.}}
\begin{center}
\begin{scriptsize}
\begin{tabular}{l l c c c c c c c l c c c c}\\
\vspace{-0.9cm}\\
\hline\vspace{-0.2cm}\\
Plasma & \multicolumn{7}{c}{$-E$ for two-electron ions}&&\multicolumn{5}{c}{$-E~$ for one-electron ions}\\
  \cline{2-8}\cline{10-14}\vspace{-0.2cm}\\
 density& R & $1s^2$ & $1s2s$ & $1s3s$ & $1s2p$ & $1s3p$ & $1s4p$ && R & $E_{1s}$ & $E_{2s}$ &$ E_{2p}$& $E_{3p}$\\
 \hline\vspace{-0.2cm}\\
            &$Ar^{16+}$&&&&&&&&$Ar^{17+}$&&\\
3.22(21) & 20.0  & 310.507 205 &196.041 306 &175.755 299 & 195.578 034 & 175.620 951 & 168.619 605 & & 20.408& 160.750 518 & 39.250 638 & 39.250 583 & 16.751 038 \\
2.58(22) & 10.0  & 308.107 341 &193.642 370 &173.360 319 & 193.178 832 & 173.225 396 & 166.234 270 & & 10.204 & 159.501 078 & 38.002 041 & 38.001 721 & 15.505 423\\
2.06(23) & 5.0   & 303.308 425 & 188.850 886 & 168.600 563 & 188.385 210 & 168.461 043 & 161.553 149 && 5.102 & 157.002 621 & 35.510 327 & 35.507 887 & 13.037 578 \\
9.55(23) & 3.0   &&&&&                                                                   & 155.654 881 &&3.061 & 153.672 785 & 32.208 489 & 32.196 858 & 9.835 118\\
2.11(24)&2.3&&&&&&151.183 198&& 2.347 &151.140 923 &29.720 258&&\\
2.96(24) & 2.057 & 289.590 037 &              &              & 174.748 078 & 155.284 420 & 148.653 416 &&2.099  & 149.859 811 & 28.470 821 & 28.436 728 & 6.367 203\\
3.22(24) & 2.0   & 288.926 557 & 174.593 816 & 154.885 258 & 174.092 315 & 154.670 904 & 147.924 331 & &2.041 & 149.514 344 & 28.135 153 & 28.099 008 & 6.066 598\\
7.64(24) & 1.5   &&                            & 147.515 430 &              & 147.294 212 & 137.296 178 &&1.531 & 145.362 038 & 24.149 851 & 24.065 320 & 2.548 088 \\
1.22(25)&&&&&&&&&1.309&&&21.374 828&0.027 313\\
1.49(25) & 1.2   &&                            & 139.209 711 &&                         & 123.092 431 && 1.224 & 141.218 060 & 20.283 441 & 20.102 175\\
2.58(25) & 1.0   & 265.062 279 & 151.634 784 & 129.063 495 & 150.887 572 & 129.848 525 & 105.387 057 && 1.020 & 137.084 181 & 16.548 570 & 16.250 296 \\
5.03(25) & 0.8   &&                            & 109.499 582 &&                            & 72.434 154&& 0.816 & 130.907 580 & 11.023 329 & 10.593 320 \\
7.51(25) & 0.7   & 244.788 710 & 140.318 457 & 92.135 024  & 131.230 710  & 97.682 171 & 44.281 500 && 0.714 & 126.516 643 & 6.834 684 & 6.481 745 \\
9.39(25) & 0.65  &&&&&                                                              & 25.066 379 &&0.663 & 123.824 592 & 4.030 093 & 3.846 749\\
1.13(26) & 0.61  &&&&&                                                              & 6.173 887  & &0.622 & 121.360 487 & 1.249 801 & 1.311 525\\
1.18(26) & 0.6014&&&&&                                       & 1.639 257 && 0.614 & 120.788 926 & 0.571 762 & 0.762 695\\
1.19(26) & 0.6   &&                            & 64.536 211 &&&&&0.612 & 120.694 413 & 0.458 366 & 0.622 378  \\
1.201(26)& 0.5987&&&&&                          & 0.228 118 & &0.6109 & 120.606 199 & 0.352 195 & 0.544 660\\
1.203(26)&0.59843&&&&&                                      & 0.067 636 && 0.6106 & 120.587 871 & 0.330 097 & 0.523 396\\
1.24(26)&&&&&&&&&0.604&&&0.048 119&\\
2.06(26) & 0.5   & 218.150 195 & 99.801 076  & 17.342 729 & 102.188 104  & 37.003 115 &&&0.510 & 112.615 414 &\\
2.19(26) & 0.49  &&                            & 10.878 182 &&&&& 0.500 &111.632 327 & \\
2.33(26) & 0.48  &&                            & 3.985 508 &&&&& 0.4898 &110.609 857 &\\
2.344(26)& 0.4791  &&                          & 3.342 920 &&& && 0.4889 &110.515 854 &\\
2.345(26)& 0.47902 &&                          & 3.285 614 &&& & &0.4888 &110.507 440 &\\
2.41(26) &0.47455&&                            & 0.035 558 &&& && 0.4842 &110.035 148 &\\
2.83(26) & 0.45    &&&&                         & 8.128 548   &&& 0.4592 &107.280 138 &\\
3.053(26)& 0.4387  &&&&                          & 0.218 634   &&& 0.448 &105.912 248 &\\ 
3.059(26)& 0.43843 &&&&                  & 0.022 169   &&    &  0.447 & 105.878 796 &\\
4.03(26) & 0.4     &            & 63.058 578  &             & 72.434 796  &&  & &0.408& 100.662 918 &\\
6.01(26) & 0.35    &            & 30.887 328  &&&&& & 0.357 & 92.198 308 &\\
7.87(26) & 0.32    &            & 3.274 140  &&&&&& 0.326 &85.821 161 &\\
8.01(26) & 0.318   &            & 1.126 899  &&&&& & 0.3245  & 85.350 316 &\\
8.04(26) & 0.3176  &            & 0.692 178  &&&&& & 0.3241 &85.255 417 &\\
8.05(26) & 0.31751 &            & 0.594 121  &&&&& & 0.3240 &85.233 983 &\\
8.09(26) & 0.31698 &            & 0.014 841  &&&&&&  0.323 &85.107 902 &\\
9.55(26) &0.3      &156.809 245 &&&&&&& 0.306 & 80.799 235 &\\           
1.08(27) &0.288    &&&&0.703 578&&&& 0.294 & 77.402 624 &\\
1.09(27) &0.2874   &&&&0.09 253 &&& & 0.293 & 77.224 113 &\\
3.22(27) &0.2      &64.368 049 &&&&&& & 0.204 &34.699 257 &\\
4.42(27) &0.18     &25.307 474 &&&&&&& 0.184  &15.611 535 &\\
5.15(27) &0.171    & 2.741 398 &&&&&&& 0.1745 & 4.631 691 &\\
5.24(27) &0.1701   & 0.269 467 &&&&&&& 0.1736 & 3.429 550 &\\
5.25(27) &0.17001  & 0.019 935 &&&&&&& 0.1735 & 3.308 080 &\\
 \hline\hline
\end{tabular}
\end{scriptsize}
\end{center}
\end{sidewaystable}

\begin{table}[!th]
\begin{center}
\caption {\rm {Critical plasma electron densities after which spectral lines of hydrogen-like and helium-like \textit{Al} disappear. Densities are obtained from IS radii according to both SP model [eq. (4)] and EK model[eq. (5)] for IPDs. The notation $x(y)$ indicates $x \times 10^{y}$.}}
\begin{tabular}{c c c c l l}\\
  \hline\hline\vspace{-0.3cm}\\
 &  & \multicolumn{4}{c}{Critical plasma electron density (per c.c.)}\\
  \cline{3-6}\vspace{-0.2cm}\\
 Z  & Spectral &\multicolumn{2}{c}{Present results} &\multicolumn{2}{c}{Other results}\\
 \cline{3-4} \cline{5-6}\vspace{-0.2cm}\\
  & line& SP model$^a$& EK model$^b$ & Experiment & \multicolumn{1}{c}{Theory} \\
 \hline\vspace{-0.2cm}\\
 6 & $Ly_\alpha$ & 1.85(24) & 1.19(24)& &\\
   & $Ly_\beta$  & 8.32(22) & 2.77(22)\\
   & $He_\alpha$ &8.05(23) & 1.61(23)\\
   & $He_\beta$  & 5.15(22) & 1.03(22)\\
   & $He_\gamma$ & 1.88(22) & 3.76(21)\\
    \hline\vspace{-0.2cm}\\
13 & $Ly_\alpha$ & 3.52(25) & 2.71(24) & &\\
   & $Ly_\beta$  & 3.31(24) & 2.55(23) & 2.2(24)$^c$ & $2.93$(24)$^d$\\
   &&&&                                              & $2.64(24)-3.3(24)^e$\\
   & $He_\alpha$ & 2.43(25) & 2.53(24) &\\
   & $He_\beta$  & 2.21(24) & 1.84(23) & 2.2(24)$^c$ & $2.44$(24)$^d$\\
   &&&&&$1.98(24)-2.64(24)^e$\\
   & $He_\gamma$ & 5.00(23) & 4.16(22) &&$6.60(22)-1.32(23)^e$\\
    \hline\vspace{-0.2cm}\\
18 & $Ly_\alpha$ & 1.24(26) & 6.90(25)\\
   & $Ly_\beta$  & 1.22(25) & 6.78(23)\\
   & $He_\alpha$ & 7.51(25) & 4.42(25)\\
   & $He_\beta$  & 7.64(24) & 4.49(23)\\
   & $He_\gamma$ & 2.11(24) & 1.25(23)\\
    \hline\hline\vspace{-0.2cm}
\end{tabular}
\end{center}\vspace{-0.4cm}
\begin{footnotesize}
~~~~~~~~~~~~~~~~~~$a$: Ref. \cite{ste28}; $b$: Ref. \cite{ek}; $c$: Ref. \cite{hoa}; $d$: Ref. \cite{son}; $e$: Ref. \cite{pre}.       
\end{footnotesize}
\end{table}

\begin{table}[!th]
\begin{center}
\caption {\rm {Transition energies (in \AA) in strongly coupled plasma environment. Densities are calculated according eq. (4). Conversion factors : $(\Delta E)_{\AA}$ = 12395/$(\Delta E)_{eV}$ and 1 $a.u.$ of energy = 27.21138 $eV$. The notation $x(y)$ indicates $x \times 10^{y}$.}}
\begin{tabular}{c l c r c c c c}\\
  \hline\hline\vspace{-0.2cm}\\
 &Ion-sphere & Plasma density &\multicolumn{1}{c}{Transition} & Present & Experimental &\multicolumn{2}{c}{Other theory$^d$}\\
  \cline{7-8}\vspace{-0.2cm}\\
Ion & radius (a.u) & (per c.c.) &\multicolumn{1}{c}{scheme} & results & results & Non-rel.& Rel. \\
 \hline\vspace{-0.3cm}\\
$C^{4+}$   & 16.256 & 1.5(21)& $1s^2 \rightarrow 1s2p$ &40.266  &40.268$^a$ &40.208&40.700 \\ 
           &        & &      $\rightarrow 1s3p$ &34.979  &34.998$^a$& 34.953 & 35.325\\
           &        & &      $\rightarrow 1s4p$ &33.451  &33.469$^a$& 33.431 & 33.773\\
           &        & &      $\rightarrow 1s5p$ &32.820  &32.773$^a$& 32.800 & 33.133\\
\hline\vspace{-0.3cm}\\
$Al^{11+}$ & 2.607  & 1.0(24)& $1s^2 \rightarrow 1s2p$ & 7.777  & 7.75$^b$&  7.778 &  7.774\\
           &        & &      $\rightarrow 1s3p$ & 6.682 && 6.684 & 6.682\\
           & 3.285  & 5.0(23)& $1s^2 \rightarrow 1s2p$ &7.773 && 7.773 & 7.770\\
           &        & &      $\rightarrow 1s3p$ & 6.663   &&  6.664 & 6.661 \\
           &        & &      $\rightarrow 1s4p$ & 6.366   && &  6.385\\
           & 5.617  & 1.0(23)& $1s^2 \rightarrow 1s2p$ &7.770   &&7.769  & 7.767  \\
           &        & &      $\rightarrow 1s3p$ &6.648   & 6.63$^b$& 6.648 &6.646  \\
           &        & &      $\rightarrow 1s4p$ &6.334   & 6.31$^b$& 6.334 &6.332  \\
           &        & &      $\rightarrow 1s5p$ &6.211   & 6.17$^b$&  6.205 &6.212  \\
\hline\vspace{-0.3cm}\\
$Ar^{16+}$ & 3.856 & 4.5(23) & $1s^2 \rightarrow 1s2p  $ &  3.379 & 3.363$^c$&&\\
           & 3.119 & 8.5(23) & $1s^2 \rightarrow 1s2p $  & 3.3811 & 3.364$^c$&&\\
           & 3.061 & 9.0(23) & $1s^2 \rightarrow 1s2p $  & 3.3813 & 3.365$^c$&&\\           
           &2.955  & 1.0(24) & $1s^2 \rightarrow 1s2p$ & 3.964 & 3.984$^c$ &  &  \\
           &        & &      $\rightarrow 1s3p$ & 3.382  & 3.365$^c$ &   &   \\
           &        & &      $\rightarrow 1s4p$ & 3.225  & 3.168$^c$ &   &   \\
            & 2.057  & 2.96(24)& $1s^2 \rightarrow 1s2p$ & 3.966    & & 3.964 & 3.950 \\
           &        & &      $\rightarrow 1s3p$ &3.392   & &3.382   & 3.370  \\
           &        & &      $\rightarrow 1s4p$ &3.232    & & 3.226  & 3.215  \\
           \hline\hline           
\end{tabular}
\end{center}
\begin{footnotesize}
~~~~~~~~~~~~$a$: Ref. \cite{nan23}; $b$: Ref. \cite{sae24}; $c$: Ref. \cite{woo26}; $d$: Ref. \cite{pkm27}           
\end{footnotesize}
\end{table}

\begin{table}[!th]
\begin{center}
\begin{footnotesize}
\caption {\rm {Thermodynamic pressure on the ground state of one and two-electron ions within ion-sphere. Conversion factor : 1 a.u. of pressure = 2.9421912(13) Pa. The notation $x(y)$ indicates $x \times 10^{y}$.}}
\begin{tabular}{l c c c l c c c l c c}\\
  \hline\hline\vspace{-0.2cm}\\
 Plasma&\multicolumn{2}{c}{Pressure (Pa)}&\vline & Plasma &\multicolumn{2}{c}{Pressure (Pa)} &\vline & Plasma &\multicolumn{2}{c}{Pressure (Pa)}\\
  \cline{2-3}\cline{6-7}\cline{10-11}\vspace{-0.2cm}\\
density & $C^{4+}$ & $C^{5+}$ & \vline & density & $Al^{11+}$ & $Al^{12+}$& \vline &density & $Ar^{16+}$ & $Ar^{17+}$\\
 \hline\vspace{-0.2cm}\\
8.05(20)& 0.1755(09) & 0.8778(09)&\vline & 2.21(21)& 0.4829(09) & 0.2414(10)&\vline & 3.22(21)& 0.7024(09) & 0.3512(10)\\
6.44(21)& 0.2807(10) & 0.7019(10)& \vline & 1.77(22)& 0.7725(10) & 0.1931(11)&\vline & 2.58(22)& 0.1123(11) & 0.2809(11)\\
5.15(22)& 0.4478(11) & 0.5603(11)& \vline & 5.17(22)& 0.3217(11) & 0.5630(11)&\vline & 2.06(23)& 0.1797(12) & 0.2247(12)\\
2.39(23)& 0.3432(12) & 0.2581(12)& \vline & 2.21(24)& 0.4806(13) & 0.2406(13)&\vline & 3.22(24)& 0.7007(13) & 0.350413)\\
1.57(24)& 0.4130(13) & 0.1666(13)&\vline & 1.77(25)& 0.7580(14) & 0.1899(14)&\vline & 2.58(25)& 0.1113(15) & 0.2785(14)\\
6.44(24)& 0.2779(14) & 0.6681(13)& \vline & 5.17(25) & 0.3095(15) & 0.5438(14)&\vline & 2.06(26) & 0.1729(16) & 0.2818(15)\\
1.88(25)& 0.1540(15) & 0.1794(14)& \vline & 1.41(26) & 0.1178(16) & 0.8791(14)&\vline & 3.22(27) & 0.9657(17) & 0.3184(16)\\
5.15(25)& 0.9278(15) & 0.1772(14)&\vline & 6.56(26) & 0.1228(17) & 0.1022(16)&\vline & 4.42(27)& 0.1667(18) & 0.4212(16)\\
6.21(25)& 0.1297(16) & 0.9241(14)& \vline & 1.28(27)& 0.3912(17) & 0.1489(16)&\vline & 5.15(27) & 0.2180(18) & 0.8701(16) \\
6.23(25)& 0.1304(16) & 0.1123(15)& \vline & 1.419(27)& 0.4678(17) & 0.2453(16) &\vline & 5.24(27) & 0.2241(18) & 0.8897(16) \\
6.24(25)& 0.1309(16) & 0.1164(15)& \vline & 1.424(27)& 0.4710(17) &0.2466(16) &\vline & 5.25(27)& 0.2248(18) & 0.8916(16)\\
\hline\hline\vspace{-0.2cm}\\
\end{tabular}
\end{footnotesize}
\end{center}
\end{table}

\begin{figure}
\centerline{\epsfxsize=12.0cm\epsfysize=10.0cm\epsfbox{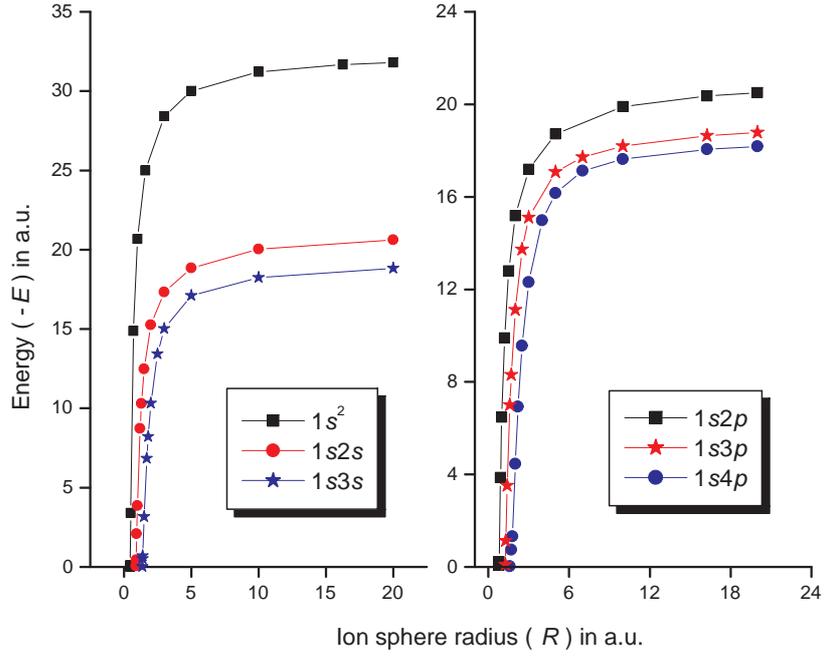}}
\caption{Variation of eigen energies ($-E$) of bound $1sns$($^1S^e$) [$n=1-3$] and $1sn'p$($^1P^o$) [$n'=2-4$] states of $C^{4+}$ with respect to ion-sphere radius $R$.}
\end{figure}

\begin{figure}
\centerline{\epsfxsize=12.0cm\epsfysize=10.0cm\epsfbox{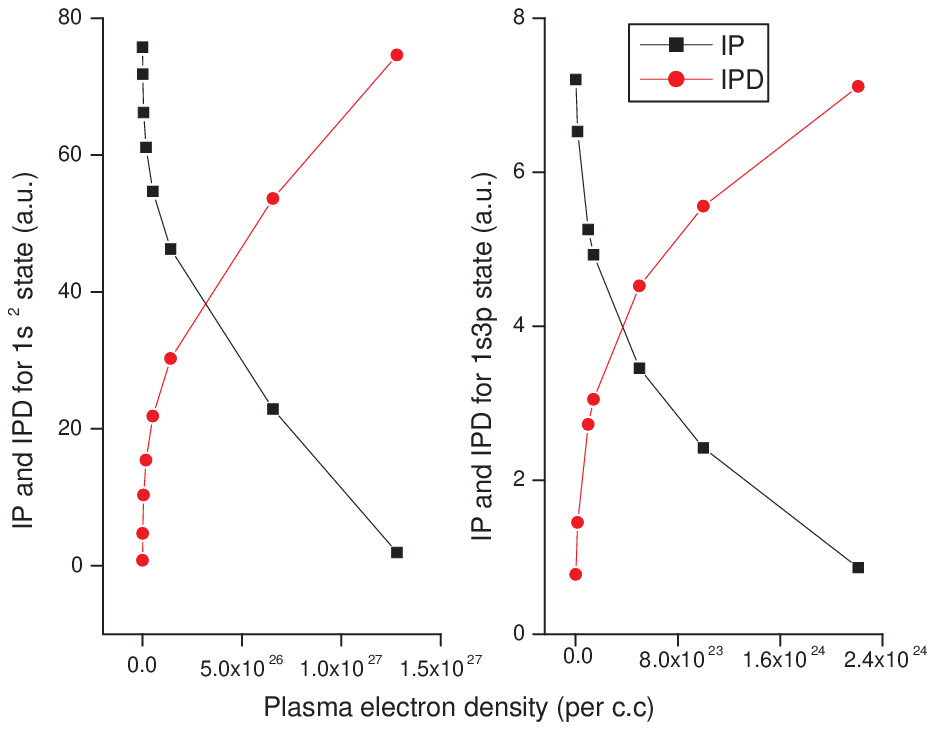}}
\caption{Variation of IP and IPD  for $1s^2$ ($^1S^e$) and $1s3p$($^1P^o$) states of $Al^{11+}$ with respect to plasma electron density.}
\end{figure}


\begin{thebibliography}{80}
\bibitem{aqc1}J. Sabin and E. Brandas (\textit{Ed.}) and prefaced by S. A. Cruz, Adv. Quantum Chem. \textbf{58} (2009).
\bibitem{jac2}P. A. Jacobs, Carboniogenic Activity of Zeolites (Amsterdam: Elsevier) (1997).
\bibitem{xu3}Y. B. Xu, M. Q. Tan, and U. Becker, Phys. Rev. Lett. \textbf{76}, 3538 (1996).
\bibitem{sil5}A.N. Sil, S. Canuto, and P. K. Mukherjee, Adv. Quantum Chem. \textbf{58}, 115 (2009) and references therein.
\bibitem{can6}S. Canuto (\textit{ed}.) Solvation Effects on Molecules and Biomolecules, Computational Methods and Applications (Berlin: Springer) (2008).
\bibitem{wal7}C.A. Walsh, J. Yuan, and L.M. Brown, Philos. Mag. B \textbf{80}, 1507 (2000).
\bibitem{kos8}M. Koskinen, M. Manninen, and S. M. Reimann, Phys. Rev. Lett. \textbf{79}, 1389 (1997).
\bibitem{cio9}J. Cioslowski and E. D. Fleischmann, J. Chem. Phys. \textbf{94}, 3730 (1991).
\bibitem{tur10}L. Turker, Int. J. Hydrogen Energy \textbf{32}, 1933 (2007).
\bibitem{gui11}T. Guillot,  Planet Space Sci. \textbf{47}, 1183 (1999).
\bibitem{sah12}J. K. Saha, S. Bhattacharyya, and T. K .Mukherjee, Int. Rev. At. Mol. Phys. \textbf{3}, 1 (2012).
\bibitem{jqs13}J. K. Saha, S. Bhattacharyya, P. K. Mukherjee, and T. K. Mukherjee, J. Quant. Spec. Rad. Trans. \textbf{111}, 675 (2010).
\bibitem{jpb14}J. K. Saha, S. Bhattacharyya, P. K. Mukherjee, and T. K .Mukherjee, J. Phys. B \textbf{42}, 245701 (2009).
\bibitem{ord16}A. F. Ordonez-Lasso, J. C. Cardona, and J. L. Sanz-Vicario, Phys. Rev. A \textbf{88}, 012702 (2013).
\bibitem{ho17}L. G. Jiao and Y. K. Ho, Phys. Rev A \textbf{87}, 052508 (2013).
\bibitem{das18}M. Das, M. Das, R. K. Chaudhuri and S. Chattopadhyay, Phys. Rev. E \textbf{85}, 042506 (2012).
\bibitem{sbz19}S. B. Zhang, J. G. Wang and R. K. Janev, Phys. Rev. Lett. \textbf{104}, 023203 (2010).
\bibitem{das20}M. Das, R. K. Chaudhuri, S. Chattopadhyay, U. S. Mahapatra, and P. K. Mukherjee, J. Phys. B \textbf{44}, 165701 (2011).
\bibitem{san15}J. P. Santos, A. M. Costa, J. P. Marques, M. C. Martins, P. Indelicato, and F. Parente, Phys. Rev. A \textbf{82}, 062516 (2010).
\bibitem{akh21}A. I. Akhiezer, I. A. Akhiezer, R. A. Polovin, A. G. Sitenko and K. N. Stepanov, Plasma Electrodynamics, Linear Response Theory (Oxford, Pergamon), Vol. 1 (1975).
\bibitem{ich22}S. Ichimaru, Rev. Mod. Phys. \textbf{54}, 1017 (1982).
\bibitem{nan23}M. Nantel, G. Ma, S. Gu, C.Y. Cote, J. Itatani, and D. Umstadter, Phys. Rev. Lett. \textbf{80}, 4442 (1998).
\bibitem{sae24}A. Saemann, K. Eidmann, I. E. Golovkin, R. C. Mancini, E. Andersson, E. Forster, and K. Witte, Phys. Rev. Lett. \textbf{82}, 4843 (1999).
\bibitem{woo26}N. C. Woolsey, B. A. Hammel, C. J. Keane, C. A. Back, J. C. Moreno, J. K. Nash, A. Calisti, C. Mosse, R. Stamm, B. Talin, A. Asfaw, L. S. Klein, and R. W. Lee, Phys. Rev. E \textbf{57}, 4650 (1998).
\bibitem{vin}S. M. Vinko \textit{et al.}, Nature Communications \textbf{6}, 6397 (2015).
\bibitem{vin1}S. M. Vinko \textit{et al.}, Nature \textbf{482}, 59 (2012).
\bibitem{ciri}O. Ciricosta \textit{et al.}, Phys. Rev. Lett. \textbf{109}, 065002 (2012).
\bibitem{cho}B. I. Cho \textit{et al.}, Phys. Rev. Lett. \textbf{109}, 245003 (2012).
\bibitem{ste28}J. C. Stewart Jr., K. D. Pyatt, Astrophys. J. \textbf{144}, 1203 (1966).
\bibitem{ek}G. Ecker and W. Kr{\"o}ll, Phys. Fluids \textbf{6}, 62 (1963).
\bibitem{pre}T. R. Preston, S. M. Vinko, O. Ciricosta, H. K. Chung, R. W. Lee, and J. S. Wark, High Energy Density Physics \textbf{9}, 258 (2013).
\bibitem{ing}D. R. Inglis and E. Teller, Astrophys. J. \textbf{90}, 439 (1939).
\bibitem{hoa}D. J. Hoarty \textit{et al.}, Phys. Rev. Lett. \textbf{110}, 265003 (2013).
\bibitem{hoa1}D. J. Hoarty \textit{et al.}, High Energy Density Physics \textbf{9}, 661 (2013).
\bibitem{son}S. K. Son, R. Thiele, Z. Jurek, B. Ziaja, and R. Santra, Phys. Rev. X \textbf{4},  031004 (2014)
\bibitem{pkm27}A. N. Sil, J. Anton, S. Fritzsche, P. K. Mukherjee, and B. Fricke, Eur. Phys. J. D \textbf{55}, 645 (2009) and references therein.
\bibitem{aqun1}N. Aquino, A. Flores-Riveros, and J. F. Rivas-Silva, Phys. Lett. A \textbf{307}, 326 (2003).
\bibitem{flo2}A. Flores-Riveros, and A. Rodriguez-Contreras, Phys. Lett. A \textbf{372}, 6175 (2008).
\bibitem{lau35}C. Laughlin, and S. I. Chu, J. Phys. A: Math. Theor. \textbf{42}, 265004 (2009).
\bibitem{mont1}A. Flores-Riveros, N. Aquino, and H. E. Montgomery Jr., Phys. Lett. A \textbf{374}, 1246 (2010).
\bibitem{scr33}S. Bhattacharyya, J. K. Saha, P. K. Mukherjee, and T. K. Mukherjee, Physica Scripta \textbf{87}, 065305 (2013).
\bibitem{mont2}H. E. Montgomery Jr., and V. I. Pupyshev, Phys. Lett. A \textbf{377}, 2880 (2013).
\bibitem{sen38}K. D. Sen, J. Chem. Phys. \textbf{122}, 194324 (2005).
\bibitem{bha31}A. K. Bhatia, and A. Temkin, Rev. Mod. Phys. \textbf{36}, 1050 (1964).
\bibitem{tkm32}T. K. Mukherjee, and P. K. Mukherjee, Phys. Rev. A \textbf{50}, 850 (1994).
\bibitem{cpl34}J. K. Saha, S. Bhattacharyya, P. K. Mukherjee and T. K. Mukherjee, Chem. Phys. Lett. \textbf{517}, 223 (2011).
\bibitem{nel36}J. A. Nelder, and R. Mead, Comput. J. \textbf{7}, 308 (1965).
\bibitem{pra37}J. K. Saha, and T. K.Mukherjee, Phys. Rev. A \textbf{80}, 022513 (2009).
\bibitem{del}E. A. Carrillo-Delgado, I. Rodriguez-Vargas, and S. J. Vlaev, PIERS Online  \textbf{5}, 137 (2009) \textit{and references therein}.
\bibitem{cap}F. Capasso, C. Sirtori, J. Faist, D. L. Sivco, S. G. Chu, and A. Y. Cho, Nature  \textbf{358}, 565 (1992).
\end{thebibliography}
\end{document}